\documentclass[12pt]{iopart}
\usepackage{iopams}  
\usepackage{graphicx}
\usepackage{color}
\usepackage{mathrsfs}
\usepackage{bm}
\usepackage{multirow}
\usepackage{booktabs}

\newcommand{\V}[1]{\bm{#1} } 

\newcommand{\Ave}[1]{\left\langle {#1} \right\rangle} 
\newcommand{\Wh}[1]{\widehat {#1}} 

\newcommand{\Extr}[1]{ \mathop{\rm Extr}_{ #1 } }

\newcommand{\mR}{\mathbb{R}}
\newcommand{\mN}{\mathbb{N}}
\newcommand{\lb}{\left(}
\newcommand{\rb}{\right)}
\newcommand{\lbb}{\left\{}
\newcommand{\rbb}{\right\}}
\newcommand{\lsb}{ \left[ }
\newcommand{\rsb}{ \right] }

\newcommand{\Req}[1]{eq.\ (\ref{eq:#1})}
\newcommand{\BReq}[1]{Eq.\ (\ref{eq:#1})}
\newcommand{\NReq}[1]{(\ref{eq:#1})}
\newcommand{\Reqs}[2]{eqs.\ (\ref{eq:#1},\ref{eq:#2})}
\newcommand{\BReqs}[2]{Eqs.\ (\ref{eq:#1},\ref{eq:#2})}

\newcommand{\Reqss}[2]{eqs.\ (\ref{eq:#1}-\ref{eq:#2})}

\newcommand{\Rfig}[1]{Fig.\ \ref{fig:#1}}
\newcommand{\Rfigs}[2]{Figs.\ \ref{fig:#1} and \ref{fig:#2}}

\newcommand{\Lfig}[1]{\label{fig:#1}}
\newcommand{\Leq}[1]{\label{eq:#1}}
\newcommand{\Rsec}[1]{sec.\ \ref{sec:#1}}
\newcommand{\NRsec}[1]{\ref{sec:#1}}
\newcommand{\Lsec}[1]{\label{sec:#1}}
\newcommand{\be}{\begin{eqnarray}}
\newcommand{\ee}{\end{eqnarray}}
\newcommand{\ba}{\begin{array}}
\newcommand{\ea}{\end{array}}
\newcommand{\no}{\nonumber}

\newcommand{\subbe}{\begin{subequations}}
\newcommand{\subee}{\end{subequations}}

\newcommand{\argmin}{\mathop{\rm arg~min}\limits}

\begin{document}

\title{Relative species abundance of replicator dynamics with sparse interactions}

\author{Tomoyuki Obuchi\dag\footnote[3]{obuchi@c.titech.ac.jp}, Yoshiyuki Kabashima\dag\footnote[4]{kaba@c.titech.ac.jp}, and Kei Tokita\ddag\footnote[5]{tokita@is.nagoya-u.ac.jp} }

\address{
\dag
Department of Mathematical and Computing Science, \\
Tokyo Institute of Technology, Yokohama 226-8502, Japan\\
\ddag
Graduate School of Information Science, \\
 Nagoya University, Nagoya 464-8601, Japan\\
}

\begin{abstract}
A theory of relative species abundance on sparsely-connected networks is presented by investigating the replicator dynamics with symmetric interactions. Sparseness of a network involves difficulty in analyzing the fixed points of the equation, and we avoid this problem by treating large self interaction $u$, which allows us to construct a perturbative expansion. Based on this perturbation, we find that the nature of the interactions is directly connected to the abundance distribution, and some characteristic behaviors, such as multiple peaks in the abundance distribution and all species coexistence at moderate values of $u$, are discovered in a wide class of the distribution of the interactions. The all species coexistence collapses at a critical value of $u$, $u_c$, and this collapsing is regarded as a phase transition. To get more quantitative information, we also construct a non-perturbative theory on random graphs based on techniques of statistical mechanics. The result shows those characteristic behaviors are sustained well even for not large $u$. For even smaller values of $u$, extinct species start to appear and the abundance distribution becomes rounded and closer to a standard functional form. Another interesting finding is the non-monotonic behavior of diversity, which quantifies the number of coexisting species, when changing the ratio of mutualistic relations $\Delta$. These results are examined by numerical simulations, and the multiple peaks in the abundance distribution are confirmed to be robust against a certain level of modifications of the problem. The numerical results also show that our theory is exact for the case without extinct species, but becomes less and less precise as the proportion of extinct species grows.    
\end{abstract}

\maketitle
\section{Introduction}
Many large-scale systems in nature, such as food webs in ecosystems and pricing systems in markets, appear as a result of evolution involving complicated interactions between components of the systems. While those complex systems are ubiquitous and thus are desired to be understood, our theoretical and experimental schemes to treat such systems are still limited. The complicacy and the overwhelming diversity in the interactions and components make it a challenging problem to control those systems theoretically and experimentally. 

A realistic approach to understand such complex systems is to capture some characteristic macroscopic patterns of those systems. In particular, let us focus on ecology in the present paper. In this discipline, one of the most accumulated areas of knowledge of such macroscopic patterns is that of relative-species-abundance (RSA) ones. The abundance of a species, defined as the number of individuals in the species relative to the total number of individuals among all the species in a focused area, is a key quantity and all the species can be indexed by it. Less-abundant species are thought to be extinction-prone, which implies it is important for nature conservation to understand the underlying mechanism of emergence of such less-abundant species. Comprehending dominating parameters of RSA patterns will enable us to effectively prioritize actions to protect nature. 

Nevertheless, knowledge concerning the mechanism of RSA patterns is still limited. Statistical descriptions of RSA patterns have been proposed over many decades~\cite{Motomura:32,Corbet:43,MacArthur:57,MacArthur:60,Preston:62a,Preston:62b,Whittaker:70,Bazzaz:75,May:75,Sugihara:80,Nee:91,Tokeshi:99}. Deeper theoretical analyses have been advanced rather recently by the aid of recent technical developments in stochastic processes~\cite{Hubbell:01,Volkov:03,Vallade:03,Alonso:04,Etienne:05,Alonso:06,Etienne:07}. These theoretical studies are mainly based on the neutral theory which is suitable for describing systems on a given trophic level only with competition, such as coral reefs and tropical rainforests. To describe more complicated ecosystems like food webs of animals, more profound treatment is required, and one of the major theories is based on the replicator dynamics (RD)~\cite{Hofbauer:98}. The RD can describe a community of $N$ species in various types of interspecies interactions, and is used in various fields~\cite{Hofbauer:98,Taylor:78,Nowak:06,Mougi:12,Eigen:79,Ohtsuki:06,Nowak:01}. Statistical-mechanical treatment has played an important role in analyzing the RD with a large number of species~\cite{Rieger:89,Diederich:89,Oliveira:00,Oliveira:01,Oliveira:02,Oliveira:03,Tokita:04,Tokita:06,Galla:06,Yoshino:08,Galla:12}. The statistical-mechanical approach provides a great help in treating large-$N$ systems, which are difficult to treat even by experimental field research or by numerical simulations. Moreover, it has a wide applicability which allows us to employ several analytical ideas invented in different disciplines and to compare various results derived in different contexts. In this paper, we also follow this line of reasoning and analyze the RD by statistical-mechanical techniques to get new insights about RSA patterns, especially focusing on the conditions when and how extinct species emerge.

Let us describe the RD here. Consider a community of $N$ species, denote the $i$th species' population as $x_i \geq 0$, and assume the total population is fixed at $N=\sum_{i=1}^{N}x_i$. Each species $i$ is driven by the corresponding fitness function $\mathcal{F}_i\lb \V{x} \rb$ 
\be
\mathcal{F}_i \lb \V{x} \rb =-\frac{1}{2}\sum_{j}K_{ij}x_{j},
\Leq{F_i}
\ee
through the following differential equation
\be
\frac{dx_{i}}{dt}=x_{i}\lb
 \mathcal{F}_i\lb \V{x} \rb 
-\frac{1}{N} \overline{\mathcal{F}} \lb\V{x} \rb  
\rb,
\Leq{RD}
\ee
where $\overline{\mathcal{F}}(\V{x})$ is the averaged fitness
\be
\overline{\mathcal{F}}=\sum_{i=1}^{N} x_{i}\mathcal{F}_i \lb \V{x} \rb.
\Leq{fitness}
\ee
This is the RD. The RD appears in various fields such as biology, sociology, and game theory. The case of the symmetric interaction $K_{ij}=K_{ji}$ displays a simple nature such that the averaged fitness becomes a Lyapunov function and thus the dynamics necessarily converge to a certain fixed point. Even with such a simple behavior, the symmetric RD is still important since it can describe several phenomena such as competitive communities for common resources in classical game theory and a certain type of selection equations in population genetics. Also, it includes a certain class of Lotka-Volterra (LV) equation with non-symmetric interactions which is a basic model in ecology~\cite{Hofbauer:98}. The symmetric $K_{ij}$ is a crucial property in conducting statistical-mechanical analysis thus we keep this as earlier studies.

We treat the interactions $K_{ij}$ as random variables instead of giving deterministic values. This randomization assumption was first introduced in the context of ecology by May~\cite{May:72}, and has been employed in many theoretical works~\cite{Rieger:89,Diederich:89,Oliveira:00,Oliveira:01,Oliveira:02,Oliveira:03,Tokita:04,Tokita:06,Galla:06,Yoshino:08,Galla:12,Allesina:12}. Although this randomization of the interactions is not necessarily realistic, considering the complicacy of the experimentally-estimated interactions among species~\cite{Berlow:99}, we expect that the randomized interactions will be a good starting point to capture the macroscopic behavior of such complicated ecosystems. 

One unsatisfied assumption in the earlier statistical-mechanical studies of the RD is that each species interacts with (almost) all other species, which is clearly unrealistic in ecology. Instead of that, we here investigate the RD with sparse interactions between the species. Thanks to this sparseness, all species coexistence naturally happens in a certain region of the parameters in our model. This is really in contrast to the fully-interacting cases~\cite{Rieger:89,Diederich:89,Oliveira:00,Oliveira:01,Tokita:04,Tokita:06,Galla:06,Yoshino:08,Galla:12}. By changing the parameters, we also observe that all species coexistence collapses and extinct species start to emerge. This change can be regarded as a phase transition. This transition is also observed in \cite{Oliveira:02,Oliveira:03} of a fully-connected model, but our model is more natural in that the species does not constitute any modular structure a priori, in contrast to the ones~\cite{Oliveira:02,Oliveira:03} which are separated into a few groups, where the species in a group take a common number of individuals. 

Another interesting property revealed by our analysis is that the abundance distribution exhibits multiple peaks in a certain region of the parameters. Correspondingly, the diversity, which quantifies the number of coexisting species, shows a non-monotonic behavior when the parameter controlling the ratio of mutualistic relations, $\Delta$, changes. These properties may be compared with multiple peaks observed in several experimental data~\cite{Dornelas:08,Gray:05,Magurran:03}. A theoretical analysis of multiple peaks was provided in~\cite{Alonso:08}, but it explicitly assumes the presence of multiple peaks in the abundance distribution. We again stress that our model does not assume any multiple peaks of the abundance distribution or any modular structure in the species a priori. Our multiple peaks come from the sparseness of the interactions and the loose discreteness of the distribution of the interactions. These assumptions can be reasonable in some realistic situations, and thus our theory will provide a considerable clue to understand such multiple peaks observed in several field dataset. 

The remainder of the present paper is as follows. In the next section, we formulate the problem as energy minimization in a physics context, and solve it by neglecting the constraint $x_i \geq 0$, which is justified if the self interaction $u$ is large enough. Further, some practical information is extracted by an perturbative expansion with respect to $u^{-1}$. In \Rsec{Boltzmann}, we reformulate this by using the Boltzmann distribution. An approximation called Gaussian approximation is introduced and shown to be equivalent to neglecting the constraint $x_i \geq 0$ in the previous section. Benefits of this formulation are additional information on the variance of each species, which can be connected to the stability of the species against fluctuations in self interactions, and the availability of some systematic analytical techniques of statistical mechanics. Employing those techniques, in \Rsec{non-perturbative} we construct a non-perturbative theory on random graphs. This enables us to obtain more detailed quantitative information of the energy, order parameters, abundance distribution, and related quantities. Some numerical simulations are also performed to compare with these theoretical results. The comparison shows our theory is exact for large $u\geq u_c$, where $u_c$ is the critical value at which the all species coexistence starts to collapse, but does not give a precise result for $u<u_c$. The last section is devoted to the conclusion.

\section{Formulation as energy minimization}
The symmetric RD converges to fixed points as stated above. We investigate the properties of those fixed points, which can be formulated as the minimization problem of the following energy function or the Hamiltonian:
\be
&&
\mathcal{H}\lb \V{x},r| J \rb
=\frac{1}{2} \sum_{i,j}K_{ij}x_i x_j -r\lb \sum_{i}x_i-N\rb
\no \\  &&
=\frac{1}{2}u\sum_{i}x_i^2- \sum_{ \Ave{i,j} }J_{ij}x_i x_j -r\lb \sum_{i}x_i-N\rb.
\Leq{Hamiltonian}
\ee
We introduce a Lagrange multiplier $r$ to hold $\sum_{i}x_i=N$, and divide the interaction matrix $K$ into the self-interacting part $K_{ii}=u>0$ and pairwise interacting one $K_{ij}=-J_{ij}\,\,(i\neq j)$. Positive and negative $J_{ij}$ represent mutualistic and competitive relations, respectively. The symbol $\sum_{\Ave{i,j}}$ represents the summation over all the interacting pairs. Although there are some local minima in general, which correspond to different fixed points and can be meaningful depending on initial conditions of the RD, we only focus on the global minimum, or the ground state. The minimizer of the Hamiltonian can be formally written as   
\be
\V{x}^{*}=
\argmin_{ \{ x_i \geq 0\}_{i=1}^{N} }
\lb
\Extr{r}
\mathcal{H} \lb \V{x},r| J_{ij} \rb
\rb
.
\Leq{extremization}
\ee
In spite of the simple appearance of \Req{extremization}, the evaluation of $\V{x}^*$ is not easy in general situations. A mathematical origin of this difficulty is the non-negativity constraint $x_{i}\geq 0$. Fortunately, in the fully-connected interaction case, this problem is not so serious since the distribution of the effective field on a site is not strongly affected by the non-negativity constraint thanks to the law of large numbers. In the present case with sparse interactions, we cannot expect the effect of the law of large numbers since the number of interactions connected to a site is not extensive, thus the distribution of the effective field should be self-consistently determined by taking into account the non-negativity constraint. Unfortunately, we could not fully resolve this problem. As seen below, we can construct a legitimate solution of the problem if there is no extinct species $x_i>0\, (\forall{i})$, but this solution just becomes a unjustified approximation after extinct species start to emerge. Our theory, however, still provides nontrivial RSA patterns well controlled by a small number of parameters, and it is enough to see the transition between the absence and presence of extinct species, which enforces the significance of the present study.

\subsection{Direct minimization in large $u$ limit}
In the limit $u\to \infty$, the corresponding solution of \Req{extremization} becomes $x_i^*=1$. This observation justifies taking a direct variation of the Hamiltonian with respect to $\V{x}$ by neglecting the non-negativity constraint if $u$ is large enough. The variational conditions with respect to $\V{x}$ and $r$ give the compact analytic forms
\be
&&
r=\frac{N}{\sum_{i,j}K^{-1}_{ij}}.
\Leq{r-extremization}
\\
&&
x_{i}^*=r\sum_{j}K^{-1}_{ij}=N\frac{\sum_{j}K^{-1}_{ij}}{\sum_{i,j}K^{-1}_{ij}}.
\Leq{x-general}
\ee
To obtain lucid information from \Req{x-general}, we investigate the perturbation with respect to $u^{-1}$ below.

\subsubsection{Perturbative expansion}
\Lsec{perturbation}
We can expand $K^{-1}$ as follows:
\be
K^{-1}=\lb u I - J \rb^{-1}=\frac{1}{u}\sum_{p=0}^{\infty} u^{-p}J^{p}.
\ee
Insertion of this into \Req{x-general} reads
\be
x_{i}=\frac{
1+u^{-1}\sum_{j}J_{ij}+u^{-2}\sum_{j,k}J_{ij}J_{jk}+\cdots
}{
1+u^{-1}\frac{1}{N}\sum_{i,j}J_{ij}+u^{-2}\frac{1}{N}\sum_{i,j,k}J_{ij}J_{jk}+\cdots
}.
\Leq{x-series}
\ee
From this equation, we can find several interesting behaviors. For example, if the interaction is generated from a distribution consisting of disconnected multiple supports, the support of the distribution of $x_i$ also consists of disconnected regions, leading to a discrete shape of the abundance distribution. For the purpose of clear discussion, hereafter we assume that the interaction is drawn from the following distribution
\be
P(J_{ij})=\frac{1+\Delta}{2}\delta(J_{ij}-1)+\frac{1-\Delta}{2}\delta(J_{ij}+1),
\Leq{J-dist}
\ee
and that each site $i$ is connected to $c$ sites. Under these assumptions, we can derive the abundance distribution with a discrete nature. The abundance distribution $P(x)=(1/N)\sum_{i=1}^{N}\delta(x-x_i)$ is exactly the same as the probability distribution of $x$. Thus up to the first order of $u^{-1}$, \Req{x-series} gives $x_i\approx 1+u^{-1}\lb \sum_{j}J_{ij}-(1/N)\sum_{i,j}J_{ij}\rb$, which directly yields
\be
P(x)=
\sum_{k_1=0}^{c}
\lb 
\begin{array}{cc}
  c    \\
  k_{1}      
\end{array}
\rb
\lb \frac{1+\Delta}{2} \rb^{c-k_1}\lb \frac{1-\Delta}{2} \rb^{k_1}
\delta\lb x- \lb 1+\frac{c-2k_1 -c\Delta}{u} \rb \rb.
\Leq{Px-first}
\ee
We put $(1/N)\sum_{i,j}J_{ij}=c\Delta$ justified by the law of large numbers. \BReq{Px-first} presumably gives a simple explanation about abundance distributions with multiple peaks which are actually observed in some experiments~\cite{Dornelas:08,Gray:05,Magurran:03}. The only assumptions here are the discreteness of interactions and the largeness of the self interaction or productivity $u$ in the communities. This discreteness is relatively robust even if the higher order terms of $u^{-1}$ are taken into account, which supports the plausibility of this mechanism in real biological situations. Another interesting, and a little counter-intuitive, property of \Req{Px-first} is the dependence on $\Delta$. Larger $\Delta$ provides more mutualistic relations as seen in \Req{J-dist}, but the resultant abundance distribution \NReq{Px-first} is more biased to smaller values of $x$, which is clear in the lowest value of $x$,  $x_{min}=1-c(1+\Delta)/u$, of \Req{Px-first}. This implies mutualistic communities tend to produce extinct species more easily than competitive communities, in the sense that extinct species start to appear even at larger productivity $u$. An approximation of transition point $u_{c}$, at which extinct species start to emerge, can be obtained by equating $x_{min}=0$, leading to $u_c=c(1+\Delta)$ in the first order approximation. Higher order approximations are also obtained in a similar way. Up to the second order approximation, the topology of the network does not affect the result, and a clear discussion is possible. The approximation of $u_c$ in that order is shown in \Rfig{PD2nd}. 
\begin{figure}[htbp]
\begin{center}
  \includegraphics[width=0.45\columnwidth]{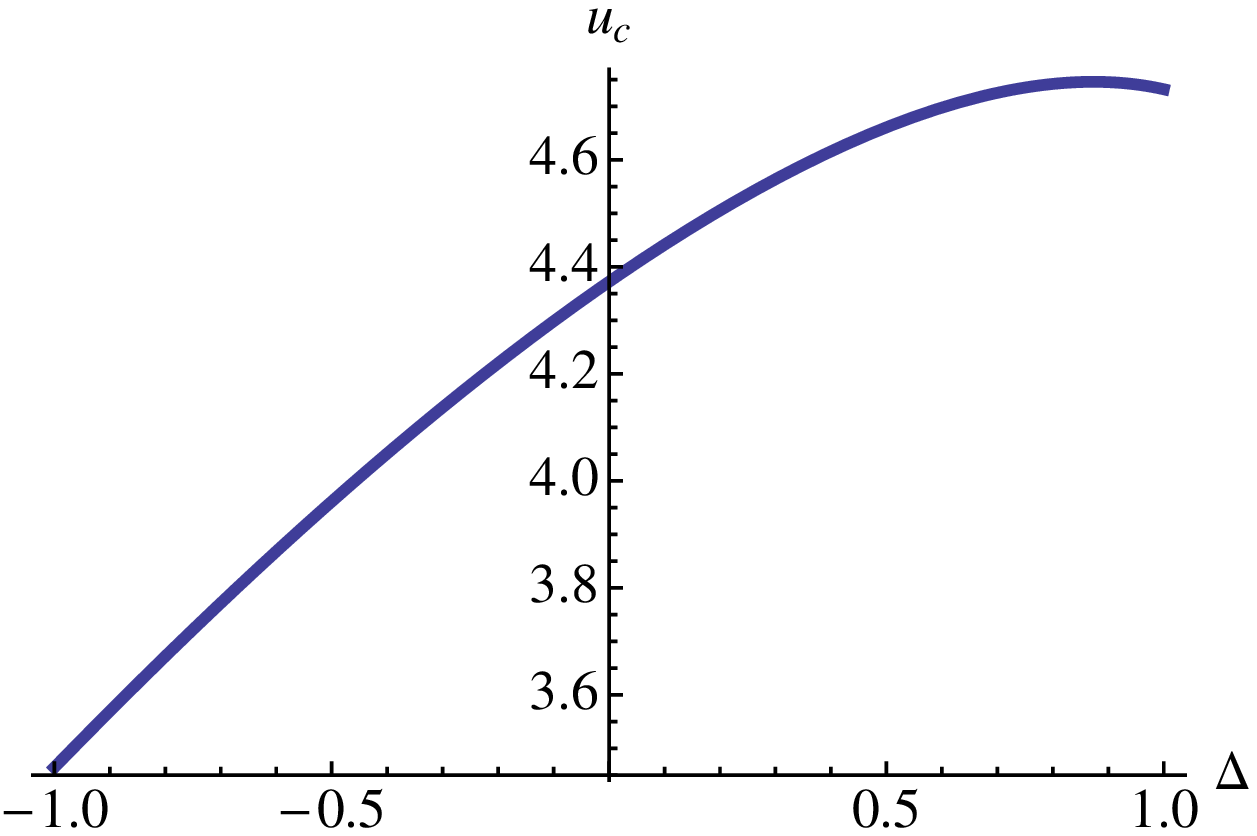}
  \includegraphics[width=0.45\columnwidth]{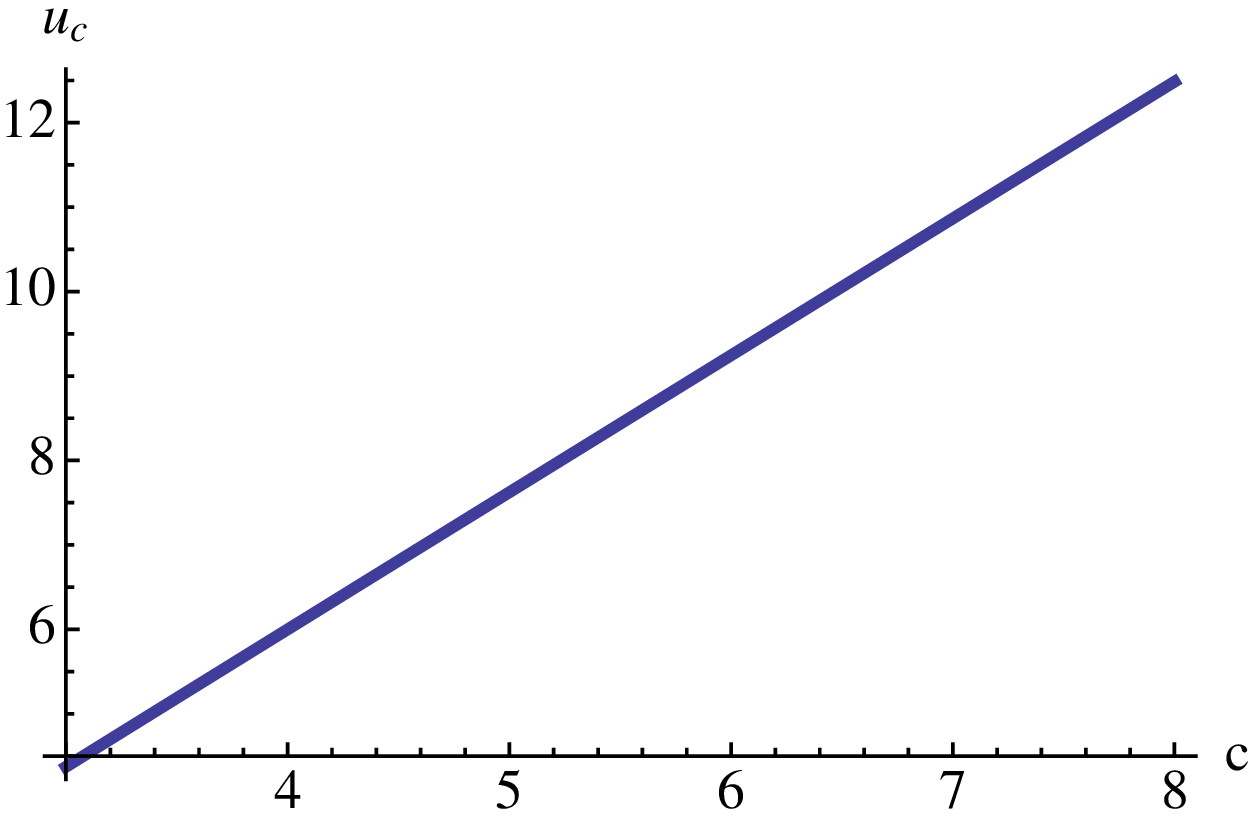}
  \caption{The values of transition point $u_{c}$ approximated by the second order expansion with respect to $u^{-1}$. The left panel is against $\Delta$ for $c=3$ and the right one is against $c$ for $\Delta=0$.}
\Lfig{PD2nd}
\end{center}
\end{figure}
From the right panel of \Rfig{PD2nd}, we can see the transition point diverges as $c$ increases, thus in the fully-connected limit extinct species always exist, which accords with the earlier analyses. 

The upper bound, or the worst-case value, of the transition point $u_{c}$ can be evaluated without truncation. The upper bound of each term in the expansion is evaluated as
\be
\left| \sum_{j_1,j_2,\cdots,j_{p}}J_{ij_1}J_{j_1j_2}\cdots J_{j_{p-1}j_p} \right| \leq c^{p}.
\ee
Then, the numerator of \Req{x-series} is bounded from below as
\be
1+u^{-1}\sum_{j}J_{ij}+u^{-2}\sum_{j,k}J_{ij}J_{jk}+\cdots
\geq 1-\sum_{p=1}^{\infty}\lb \frac{c}{u} \rb^p=\frac{u-2c}{u-c}.
\Leq{u_c-bound}
\ee
Hence, at least $u$ is larger than $2c$, and there is no extinct species irrespective of the topology of the network. This gives a firm basis of the large $u$ expansion we have investigated so far, and may provide a guiding value of self interaction in designing stable chemical reaction networks or social games.

To obtain further information other than the above bound of the transition point, we need to control the higher order terms of $u^{-1}$ in a different way. For this purpose, we construct the Boltzmann distribution of the Hamiltonian \NReq{Hamiltonian} and use a systematic approximation which becomes exact on networks without loops and when the self interaction is large enough, in the following sections.

\section{Boltzmann distribution and Gaussian approximation}\Lsec{Boltzmann}
Let us introduce the partition function $Z$ based on the standard prescription of statistical mechanics:
\be
Z(\beta, J)=
\int_{-i\infty}^{i\infty}dr \int_{0}^{\infty}\prod_{i}dx_i e^{-\beta \mathcal{H}(\V{x},r|J)}
\ee
The integration with respect to $r$ corresponds to the extremization condition with respect to $r$ in \Req{extremization}. The Boltzmann distribution is defined by $P(\V{x}|\beta,J)=\int_{-i\infty}^{i\infty}dr e^{-\beta \mathcal{H}}/Z$. In the $\beta \to \infty$ limit, the minimum-energy configuration of $\V{x}$, the ground state, is emphasized and dominates the integrations, to reproduce the result based on \Req{extremization}. If the self interaction $u$ is large enough, the ground state will be $x_{i}\approx 1$. Actually, if we neglect the pairwise interactions, the partition function can be transformed as
\be
Z=
\int_{-i\infty}^{i\infty}dr~
e^{N\beta\lb \frac{1}{2}\frac{r^2}{u}-r  \rb}
\lb \int_{0}^{\infty} dx_i~e^{-\frac{1}{2}\beta u\lb x_i- \frac{r}{ u}\rb^2}\rb^N.
\ee
In the limit $\beta \to \infty$, the saddle-point method gives the exact result and $r^*=u$ and $x_{i}^{*}=r/u=1$. The integration with respect to $x_i$ is completely dominated by around $x_i^{*}=1$, which tells that we may extend the integration region from $\lsb 0 ; \infty \rsb$ to $\lsb -\infty ; \infty \rsb$ in the limit $\beta \to \infty$. This will be the case even if the interactions exist but are small enough compared to the self interaction $u$. This corresponds to the approximation used in deriving \Req{extremization}. Let us call this approximation Gaussian approximation. 

To directly see the accordance between the Gaussian approximation and \Req{extremization}, it is appropriate to calculate the Gibbs free energy $G$  instead of the Helmholtz free energy $F=-(1/\beta )\ln Z$. This is defined as
\be
-\beta G\lb r, \V{\mu},\V{v} |\beta ,J\rb=
\beta r\lb \sum_{i}\mu_i-N\rb -\beta G_{G}\lb \V{\mu},\V{v} |\beta ,J\rb
,
\ee
where $G_{G}\lb \V{\mu},\V{v} |\beta ,J\rb$ is the purely Gaussian part of the free energy 
\be
&&
-\beta G_{G}\lb \V{\mu},\V{v} |\beta ,J\rb
\no \\ &&
\equiv
\Extr{\V{u},\V{t}}\lbb
\ln 
\int_{-\infty}^{\infty}\prod_{i}dx_i
~e^{
-\beta \mathcal{H}(\V{x},0|J)
-\frac{1}{2}\sum_{i=1}^{N}t_i\lb (x_i-\mu_i)^2-v_i \rb
+\sum_{i=1}^{N}u_i (x_i-\mu_i)
}
\rbb.
\ee
The parameters $\V{\mu}$ and $\V{v}$ represent the first and the second moments \be
\Ave{x_i}=\mu_i,\,\, \Ave{(x_i-\mu_i)^2}=v_i,
\Leq{moment}
\ee
where $\Ave{\cdots}$ denotes the average over the Boltzmann distribution. The parameters $\V{u}$ and $\V{t}$ are Lagrange multipliers to hold \Req{moment}. The condition $\sum_{i}x_i=N$ is imposed on the first moments $\V{\mu}$ for simplicity. Now, thanks to the Gaussian approximation, the integration with respect to $\V{x}$ is easy, even though the interactions exist. The result is
\be
&&
\hspace{-20mm}
-\beta G_{G}\lb \V{\mu},\V{v} \rb=
\frac{N}{2}\ln 2\pi
\no \\
&&
\hspace{-20mm}
+\Extr{\V{u},\V{t}}
\Biggl\{ 
 -\frac{1}{2}\Tr \ln \hat{K}
+\frac{1}{2 }\sum_{i,j}(u_{i}+t_{i}\mu_i)\hat{K}^{-1}_{ij}(u_{j}+t_{j}\mu_j)
-\sum_{i} u_i\mu_i 
-\frac{1}{2}\sum_{i}t_i(\mu_i^2-v_i) 
\Biggr\},
\Leq{G-gauss}
\ee
where 
\be
\hat{K}=\beta K+t_i\delta_{ij}.
\ee
The extremization condition with respect to $\V{u}$ yields
\be
u_{i}+t_i\mu_i=\sum_{j=1}^{N}\hat{K}_{ij}\mu_{j}
\Rightarrow
u_{i}=\beta \sum_{j=1}^{N}K_{ij}\mu_{j}.
\ee
Inserting this into \Req{G-gauss} and taking the extremization of $\V{t}$, we get
\be
\hat{K}^{-1}_{ii}(\V{t})=v_{i}.
\Leq{t-extremization}
\ee
Putting the solution of this equation as $\V{t}^*=\V{t}^*(\V{v})$ and summarizing the above manipulations, we get
\be
&&
-\beta G \lb r, \V{\mu},\V{v} |\beta ,J\rb
=
\beta r \lb \sum_{i}\mu_i-N\rb 
\no \\ &&
+\frac{N}{2}\ln 2\pi -\frac{1}{2}\Tr \ln \hat{K}(\V{t}^*)
-\frac{1}{2}\beta\sum_{i} K_{ij}\mu_i\mu_j 
-\frac{1}{2}\sum_{i}t_i^*v_i.
\ee
The first and second moments are decoupled and thus are determined independently. Extremizing the Gibbs free energy with respect to $\V{\mu}$ and $r$, we get $\mu_{i}=N\sum_{j}K^{-1}_{ij}/\sum_{i,j}K^{-1}_{ij}$, which is exactly the same as  \Req{x-general}. The Gaussian approximation is thus confirmed to be equivalent to the direct minimization of the Hamiltonian, neglecting the non-negativity constraint $x_{i}\geq 0$. 

One benefit of this formulation is that additional information about the variance $\V{v}$ is naturally introduced. The extremization equation with respect to $\V{v}$ yields $t_{i}=0$. Combining this with \Req{t-extremization}, we obtain
\be
v_{i}=\frac{1}{\beta}K^{-1}_{ii}=\frac{1}{\beta u}
\lbb 
1+u^{-2}\sum_{j}J_{ij}^2
+u^{-3} \sum_{j,k}J_{ij}J_{jk}J_{ki}+\cdots
\rbb.
\Leq{variance}
\ee
Thus the variance vanishes in the limit $\beta \to \infty$ as expected, and the rate of decay is determined by $K^{-1}_{ii}$. We can interpret the variance $v_i$ as the susceptibility of the $i$th species' abundance to deviation in self interactions or in the productivity. The choice of the topology of the interacting network again does not affect the result up to the second order, and a clear tendency can be extracted. The resultant variance is $v_{i}\approx (1+u^{-2}c)/(\beta u)$ and thus is increased through interactions with other species, which implies the stability against productivity fluctuation becomes weakened by the interactions. This might be a little counter-intuitive again, since common conservationists' arguments advocate the stability of the community results from complex interactions among species. 

We are now ready to construct a non-perturbative theory based on the Boltzmann distribution and the Gaussian approximation. In the next section, we formulate the problem on a regular random graph (RRG) with a fixed connectivity $c$, for which exact treatment is possible thanks to the absence of loops in the network in the large-system limit.

\section{Non-perturbative solution on random graph}\Lsec{non-perturbative}
So far we have treated a fixed realization of the interaction network and constructed the perturbation theory for the realization. Hereafter we treat an ensemble of different realizations and study the average behavior over the ensemble. This looks seemingly different from the previous sections but they are essentially the same since a typical realization behavior accords with the averaged behavior in the thermodynamic limit thanks to the self-averaging property. 

A RRG is constructed as follows. Consider a sparse network of $N$ sites where each site has $c$ connectivity to other sites which are chosen in a completely random manner. The resultant graph has loops in general, but the typical length of the loop is known to be scaled as $O(\log N)$, and thus the loops are ignorable in the thermodynamic limit. The values of interactions are assigned randomly by the distribution \NReq{J-dist} after fixing the network structure. We give a schematic picture of an RRG with $c=3$ in \Rfig{RRG}. 
\begin{figure}[htbp]
\begin{center}
  \includegraphics[width=0.3\columnwidth]{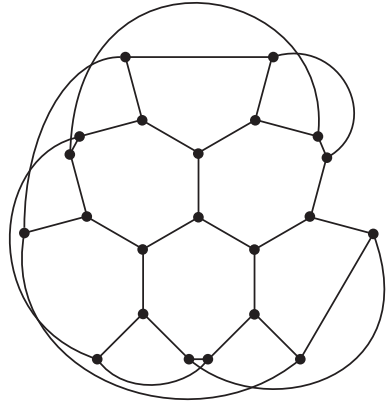}
  \caption{Schematic picture of the RRG with $c=3$ of finite size. }
\Lfig{RRG}
\end{center}
\end{figure}
Under this setting, we calculate the averaged Helmholtz free energy as
\be
-\beta \lsb F\lb \beta ,J \rb \rsb = \lsb \log Z\lb \beta ,J \rb \rsb,
\Leq{F-average}
\ee
where the square brackets $\lsb \cdots \rsb$ denote the average over the quenched randomness, i.e. both the network structure and the interaction values.
In the zero temperature limit $\beta \to \infty$, the free energy converges to the ground state energy, and the ground state exactly corresponds to \NReq{x-general}. 

In terms of the perturbative expansion \NReq{x-series}, the analysis in this section corresponds to summing up all the terms without loops. Thus, the theory on RRGs can be regarded as a non-perturbative treatment of the expansion and is known to be equivalent to the Bethe approximation.

\subsection{Analysis based on the replica and cavity methods}
Unfortunately, it is hard to take an average over the quenched randomness. This problem is circumvented by using the so-called replica method, symbolized by the following identity
\be
-\beta \lsb F\lb \beta ,J \rb \rsb =\lim_{n\to 0}\frac{1}{n} \log \lsb Z^n \rsb.
\ee 
The average of the power of the partition function, $\lsb Z^n \rsb$, is tractable if $n\in \mN$. Hence, we evaluate $\lsb Z^n \rsb$ for $n\in \mN$ and construct its analytic continuation from $n\in \mN$ to $n\in \mR$, then take the $n \to 0$ limit to finally obtain the free energy. 

We work on the Gaussian approximation. The Gaussian model on RRGs has been investigated by the replica method in some previous studies~\cite{STAT,Kabashima:12}. Therefore we do not show the detailed calculations, but just start from the explicit formula of the free energy. Readers interested in the details can see the derivation of the free energy in \NRsec{app:replica}, and refer to~\cite{Kabashima:12}. With the replica symmetry (RS), the free energy density $f=F/N$ can be expressed as
\be
&&
-\beta f=\frac{c}{2}\int d\V{\xi}_1d\V{\xi}_2 q(\V{\xi}_1)q(\V{\xi}_2) \lsb \log K_1 \rsb_{J}
-c \int d\V{\xi}d\Wh{\V{\xi}}q(\V{\xi})\Wh{q}(\Wh{\V{\xi}}) \log K_2 
 \no \\ &&
 +\int \prod_{l=1}^{c} d\Wh{\V{\xi}}_l q(\Wh{\V{\xi}}_l)\log K_{3},
 \Leq{f-RS}
\ee
where the brackets $\lsb \cdots \rsb_{J}$ denote the average over the interaction $J$ by the distribution \NReq{J-dist} appearing in $K_1$ and 
\be
&&
K_{1}=
  \int_{-\infty}^{\infty}dx_1dx_2~
    p(x_1|\V{\xi}_1)p(x_2|\V{\xi}_2) 
    e^{\beta J x_1x_2},
  \\
&&
K_2=
  \int_{-\infty}^{\infty}dx~
  p(x|\V{\xi})\Wh{p}(x|\Wh{\V{\xi}}),
  \\
  &&
K_3=  
  \int_{-\infty}^{\infty}dx~
e^{\beta r(x-1)-\frac{\beta}{2}u x^2} p(x|\Wh{\V{\xi}}_1)\cdots p(x|\Wh{\V{\xi}}_{c}).
\Leq{K_3}
\ee 
The meaning of each of these formulas is as follows. The function $p(x|\V{\xi})$ is an effective marginal distribution of a site when one neighboring site is absent, which we call cavity marginal distribution hereafter, and $\V{\xi}$ denotes the parameters characterizing the distribution. The Gaussian approximation we are employing means two parameters are enough to characterize the marginal distribution and the functional form becomes Gaussian. Referring to a previous paper~\cite{Kabashima:12}, we put $\V{\xi}=\{A,H\}$ and 
\be
p(x|\V{\xi})=p(x|A,H)=\sqrt{\frac{\beta A}{2\pi}}e^{-\frac{\beta}{2}A(x-\frac{H}{A})^2},
\Leq{p-Gauss}
\ee
and similarly
\be
\Wh{p}(x|\Wh{\V{\xi}})=\Wh{p}(x|\Wh{A},\Wh{H})=e^{\frac{\beta}{2}\Wh{A}
 \lb x^2+\frac{2\Wh{H}}{\Wh{A}}x \rb  }.
\Leq{phat-Gauss}
\ee
This $\Wh{p}(x|\Wh{\V{\xi}})$ is not normalized as a probability distribution, just for convenience in calculations. The functions $q(\V{\xi})$ and $\Wh{q}(\Wh{\V{\xi}})$ are probability distributions of the corresponding parameters. It is not possible to clarify the  functional forms, but we can derive the self-consistent equations to be satisfied by $q(\V{\xi})$ and $\Wh{q}(\Wh{\V{\xi}})$, by extremizing the free energy. This will be done after further simplifying the free energy. Specifying the functional forms of $p$ and $\Wh{p}$ enables the derivation of a more particular form of $f$. In that form, it is easy to take the $\beta \to \infty$ limit. Putting the corresponding ground state energy as $f(\beta \to \infty)=\epsilon$, we get  
\be
&&
\hspace{-20mm}
\epsilon=r-\frac{c}{2}\int dA_1 dA_2 dH_1 dH_2 ~q(A_1,H_1)q(A_2,H_2)
     \lsb 
    \frac{A_2H_1^2J^2 +2A_1A_2 H_1 H_2 J + A_1 H_2^2J^2 }{2A_1A_2(A_1A_2-J^2)}
     \rsb_{J} 
\no \\&&
\hspace{-20mm}
+c \int dA dH d\Wh{A} d\Wh{H} ~q(A,H)\Wh{q}(\Wh{A},\Wh{H})
     \lb 
    \frac{A\Wh{H}^2 +\Wh{A}H^2+2A H \Wh{H} }{2A(A-\Wh{A})}
     \rb
\no \\&&
\hspace{-20mm}
-\int \lb \prod_{l=1}^{c} d\Wh{A}_ld\Wh{H}_l~q(\Wh{A}_l,\Wh{H}_l) \rb
    \frac{(r+\sum_{l=1}^{c}\Wh{H}_{l})^2 }{ 2(u-\sum_{l=1}^{c}\Wh{A}_l  ) }.
\Leq{epsilon_RS}
\ee
Taking the variation with respect to $q$ and $\Wh{q}$, we get the following self-consistent equations 
\be
&&
\hspace{-10mm}
\Wh{q}(\Wh{A},\Wh{H})=\int dA dH q(A,H)\delta(\Wh{A}-1/A)
\lsb\delta(\Wh{H}-JH/A)\rsb_J  \Leq{F2B}
,
\\
&&
\hspace{-10mm}
q(A,H)=\int \prod_{l=1}^{c-1} d\Wh{A}_ld\Wh{H}_l \Wh{q}(\Wh{A}_l,\Wh{H}_l)
\delta\lb A-\lb u-\sum_{l=1}^{c-1}\Wh{A}_l \rb \rb
\delta\lb H-\lb r+\sum_{l=1}^{c-1}\Wh{H}_l \rb \rb.
\Leq{B2F}
\ee
Finally, taking a variation with respect to $r$, we get
\be
1=\int dA dH \frac{H}{A}Q(A,H),
\Leq{normalization}
\ee
where we define
\be
\hspace{-10mm}
Q(A,H)=\int\prod_{l=1}^{c}d\Wh{A}_l d\Wh{H}_l \Wh{q}(\Wh{A}_l,\Wh{H}_l)
\delta\lb A-\lb u-\sum_{l=1}^{c}\Wh{A}_l \rb \rb
\delta\lb H-\lb r+\sum_{l=1}^{c}\Wh{H}_l \rb \rb.
\Leq{marginal}
\ee
Solving \Reqss{F2B}{normalization} and inserting the result, we can obtain the ground-state energy.

The meaning of the parameters and functions are well interpreted by the cavity method. The cavity method, based on the spirit of the mean-field theory, approximates the problem by a batch of single-body problems. The effective marginal distribution of a site $i$ can be parameterized as 
\be
P(x_{i}|A_i,H_i)\propto e^{-\frac{1}{2}\beta A_i \lb x_i-\frac{H_i}{A_i} \rb^2},
\Leq{marginal-x}
\ee
since the model we are treating is Gaussian. The cavity method calculates the parameters $A_i$ and $H_{i}$ in a self-consistent manner. For this, we introduce the cavity marginal distribution of $i$ when one neighboring site $j$ is absent, which corresponds to \Req{p-Gauss}. Denoting the parameters of the cavity marginal distribution as $p(x_{i}|A_{i\to j},H_{i\to j})\propto e^{-\frac{1}{2}\beta A_{i\to j} \lb x_i-\frac{H_{i\to j}}{A_{i\to j}} \rb^2}$, where the parameters $A_{i\to j}$ and $H_{i\to j}$ are called cavity fields, we can calculate the marginal distribution of $j$ from the cavity marginal ones of the neighboring sites through 
\be
P(x_{j}|A_j,H_j)\propto
 \int 
\lb \prod_{i \in \partial j} dx_{i}\,p(x_{i}|A_{i\to j},H_{i\to j})\rb
e^{\beta r (x_j-1)-\frac{1}{2}\beta u x_{j}^2-\beta \sum_{i \in \partial j }J_{ij}x_{i}x_{j}}.
\ee 
where $\partial j$ denotes the set of neighboring sites of $j$. This relation directly leads to 
\be
&&
A_{j}=u-\sum_{i \in \partial j} \frac{1}{A_{i \to j}}
\equiv u-\sum_{i \in \partial j}   \Wh{A}_{i \to j}  ,
\Leq{cav2gen-A}
\\
&&
H_{j}=r+\sum_{i \in \partial j}\frac{J_{ij}H_{i\to j}}{A_{i \to j}}
\equiv r+\sum_{i \in \partial j}\Wh{H}_{i\to j},
\Leq{cav2gen-H}
\ee
where we introduce the auxiliary variables $\Wh{A}_{i\to j}$ and $\Wh{H}_{i\to j}$ which can be interpreted as effective fields on the site $j$ from a neighboring site $i$ through the interaction $J_{ij}$ and are called cavity biases. \BReqs{cav2gen-A}{cav2gen-H} are simply the arguments of the delta functions in \Req{marginal}, thus the function $Q(A,H)$ is understood as the distribution of the parameters of the genuine marginal distribution. To obtain the actual values of the cavity fields or cavity biases, we need to clarify the relation between them. This is also straightforward because the cavity fields $A_{i\to j}$ and $H_{i\to j}$ are determined by the cavity biases from the neighboring sites except for $j$, which are denoted by the symbol $\partial i \backslash j$,
\be
A_{i\to j}=u-\sum_{k\in \partial i \backslash j}\Wh{A}_{k\to i},
\\
H_{i\to j}=r+\sum_{k\in \partial i \backslash j}\Wh{H}_{k\to i}.
\ee
and the transformations from the cavity fields to biases are already given in \Reqs{cav2gen-A}{cav2gen-H}. These are simply the relations of the arguments of the delta functions in \Reqs{F2B}{B2F}. In this way, the cavity fields and biases are calculated self-consistently and the replica result is interpreted. 

In the present case where $J^2=1$ and all sites are equivalent in the sense that they have a fixed equal connectivity, the value of $A$ is unique among sites and does not fluctuate, thus we can state 
\be
q(A,H)=\delta(A-a)q(H),
\,\, 
\Wh{q}(\Wh{A},\Wh{H})=\delta(\Wh{A}-\Wh{a})\Wh{q}(\Wh{H}).
\ee
Thanks to this simplicity, the ground-state energy becomes
\be
\hspace{-20mm}
\epsilon=r-\frac{c}{2}\frac{m_2+a m_1^2 \Delta }{a(a^2-1)}
+c\frac{a\Wh{m}_2+\Wh{a}m_2+2am_1\Wh{m}_1}{ 2a(a-\Wh{a}) }
-\frac{r^2+2cr\Wh{m}_1+c\Wh{m}_2+c(c-1)\Wh{m}_1^2}{ 2(u-c\Wh{a}) },
\ee
where we state
\be
m_k=\int dH q(H)H^{k},\,\,
\Wh{m}_k=\int d\Wh{H} q(\Wh{H})\Wh{H}^{k}.
\ee
Thus, the full information of $q(H)$ is not needed to calculate the ground-state energy. The extremization conditions of all the variational parameters yield simple algebraic equations. The solution gives
\be
&&
\Wh{a}=\frac{u - \sqrt{u^2-4(c-1)}}{2(c-1)},
\Leq{ahat}
\\
&&
a=\Wh{a}^{-1}, 
\Leq{a}
\\
&&
\Wh{m}_1=\frac{u-c\Wh{a}}{1+\Wh{a}\Delta}\Wh{a}\Delta,
\Leq{m_1hat}
\\
&&
m_1=\frac{u-c\Wh{a}}{1+\Wh{a}\Delta},
\Leq{m_1}
\\
&&
r=( u-c\Wh{a} )\lb 1- \frac{ c\Wh{a}\Delta }{ 1+\Wh{a}\Delta } \rb,
\\
&&
\Wh{m}_2=\Wh{a}^2
\frac{(1 - \Wh{a}^2 (c-1) \Delta^2) (u-c\Wh{a} )^2}
{(1- (c-1)\Wh{a}^2) (1 + \Wh{a} \Delta)^2},
 \Leq{m_2hat}
\\
&&
m_2=\frac{(1 - \Wh{a}^2 (c-1) \Delta^2) (u-c\Wh{a} )^2}
{(1- (c-1)\Wh{a}^2) (1 + \Wh{a} \Delta)^2}.
\Leq{m_2}
\ee
Using these relations, the ground-state energy is much simplified as
\be
\epsilon=\frac{1}{2}r.
\ee

\subsubsection{Behavior of order parameters}
We here summarize the behavior of order parameters and some related quantities. 


The marginal distribution is Gaussian, as shown in \Req{marginal-x}. We are interested in the coefficient of the quadratic term, $A_i=A$ in \Req{marginal-x}, since it is connected to the variance $v=1/(\beta A)$ in \Req{variance} and is related to a susceptibility of abundance against deviation in the self interaction, as explained in \Rsec{Boltzmann}. According to \Req{cav2gen-A}, this is simply $A=u-c\hat{a}$, and we plot it against $u$ in the left panel of \Rfig{width} for $c=3$.
\begin{figure}[htbp]
\begin{center}
  \includegraphics[width=0.45\columnwidth]{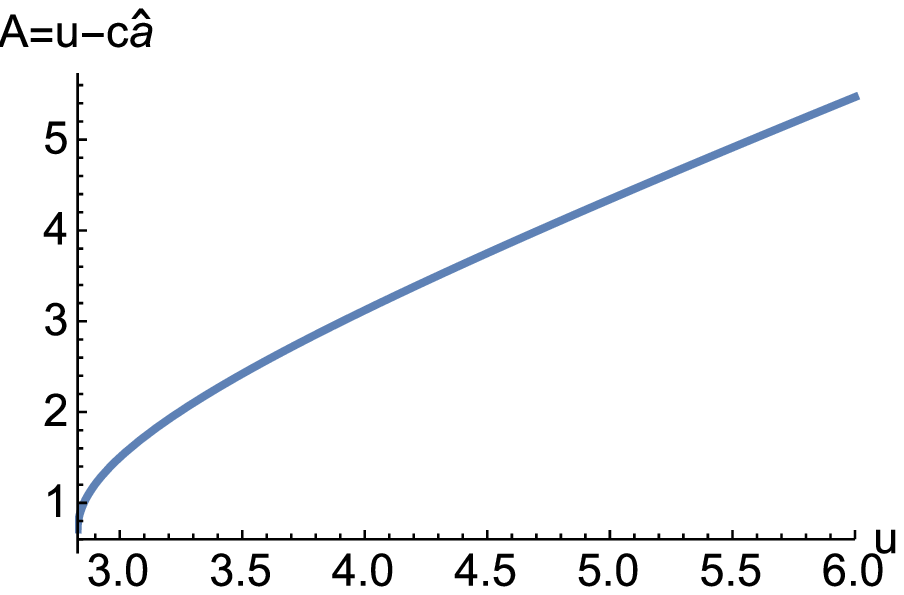}
  \hspace{2mm}
  \includegraphics[width=0.45\columnwidth]{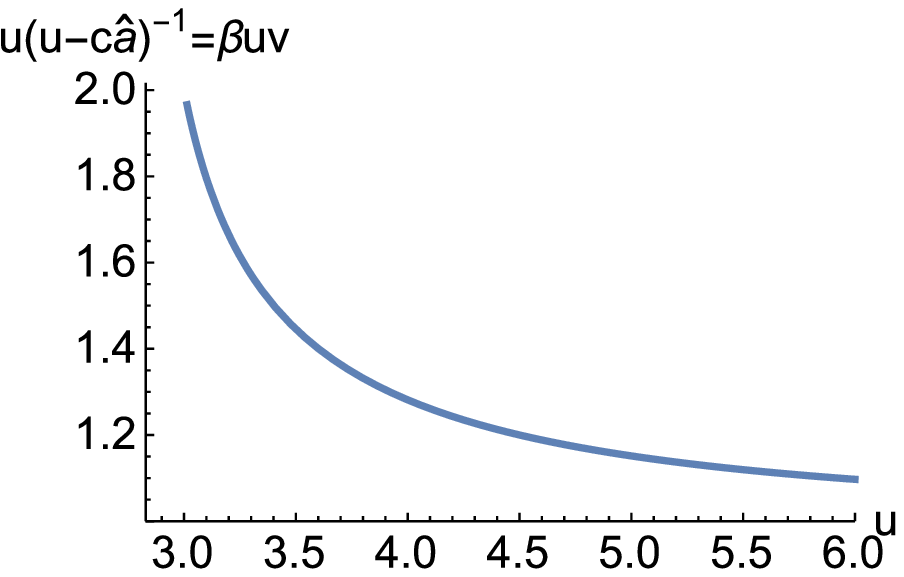}
\caption{The quadratic coefficient $A=u-c\hat{a}$ of the marginal distribution \NReq{marginal-x} (left) and its ratio to the one in the no-interaction case, $u/(u-c\hat{a})=\beta u v$ (right) where $v$ is the variance corresponding to \Req{variance}, are plotted against the self interaction $u$ for $c=3$.}
\Lfig{width}
\end{center}
\end{figure}
As seen from this panel, the positivity of the quadratic coefficient is well maintained, which is in contrast to the similar problem in the context of the first eigenvalue problem~\cite{Kabashima:12}. To quantify the effect on the quadratic coefficient by the interaction, we plot the ratio $\beta u v=u/A=u/(u-c\hat{a})$, which is unity if there is no interaction, in the right panel of the same figure. As we can see, the ratio $\beta u v$ is always larger than unity, meaning that the interactions increases the variance $v$ and thus the stability becomes weakened, which accords with the perturbation result in \Rsec{perturbation}.

The first moment $m_1$ is an increasing function of $u$ but a decreasing function of $\Delta$, and the ground-state energy $\epsilon=r/2$ as well. As examples, we plot them for $c=3$ in \Rfigs{m_1}{epsilon}  
\begin{figure}[htbp]
\begin{center}
  \includegraphics[width=0.45\columnwidth]{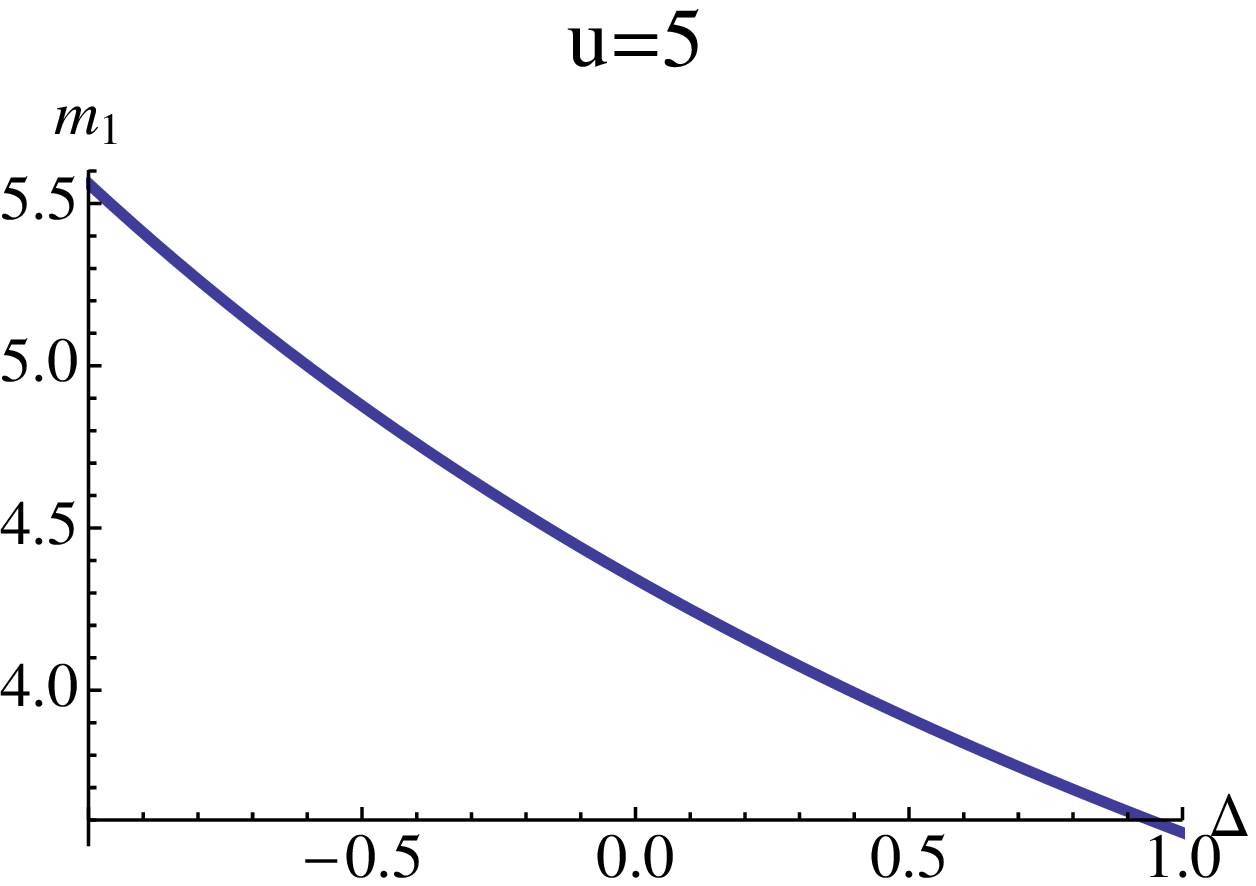}
  \hspace{2mm}
  \includegraphics[width=0.45\columnwidth]{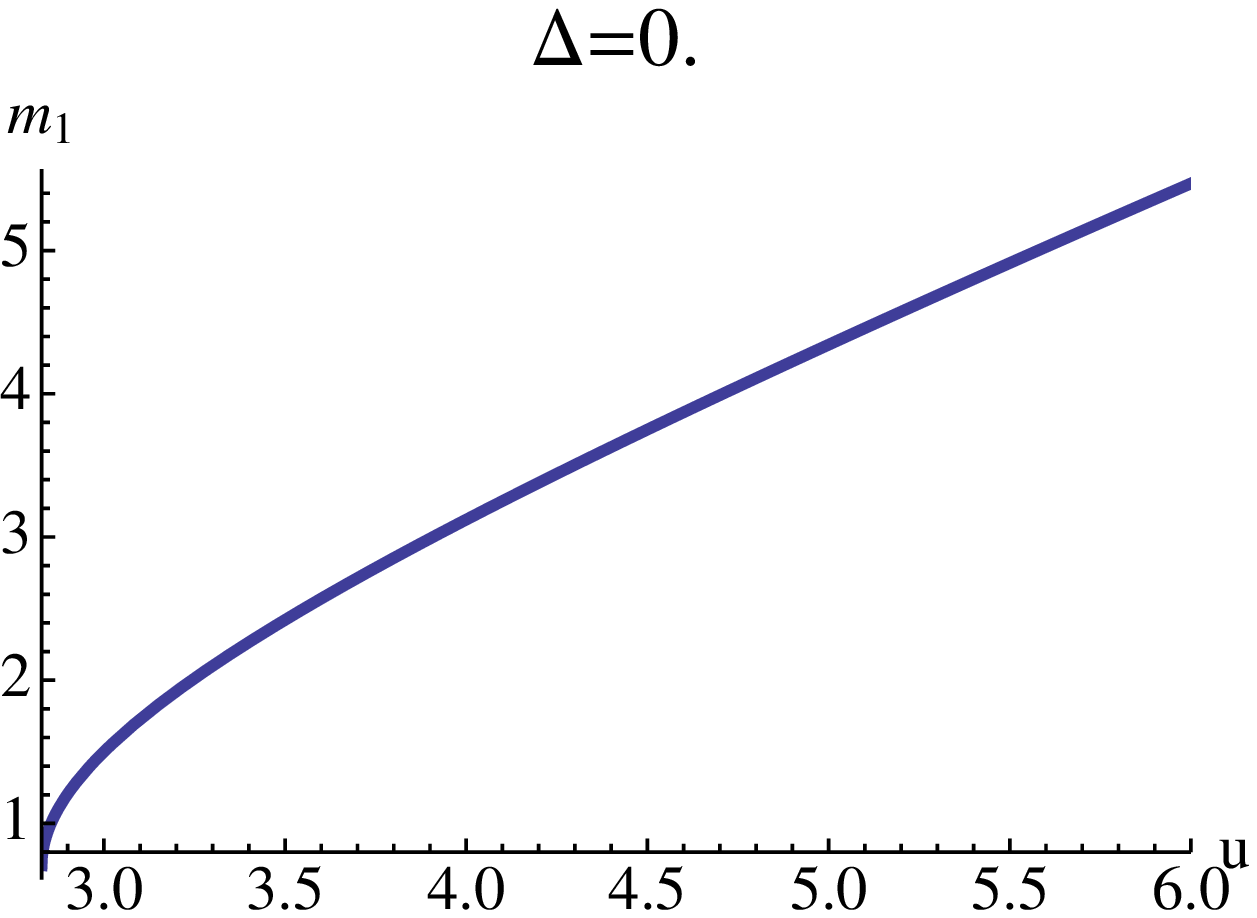}
\caption{The first moment $m_1$ for $c=3$ is plotted against $\Delta$ for $u=5$ (left) and against $u$ for $\Delta=0$ (right).}
\Lfig{m_1}
\end{center}
\end{figure}
\begin{figure}[htbp]
\begin{center}
  \includegraphics[width=0.45\columnwidth]{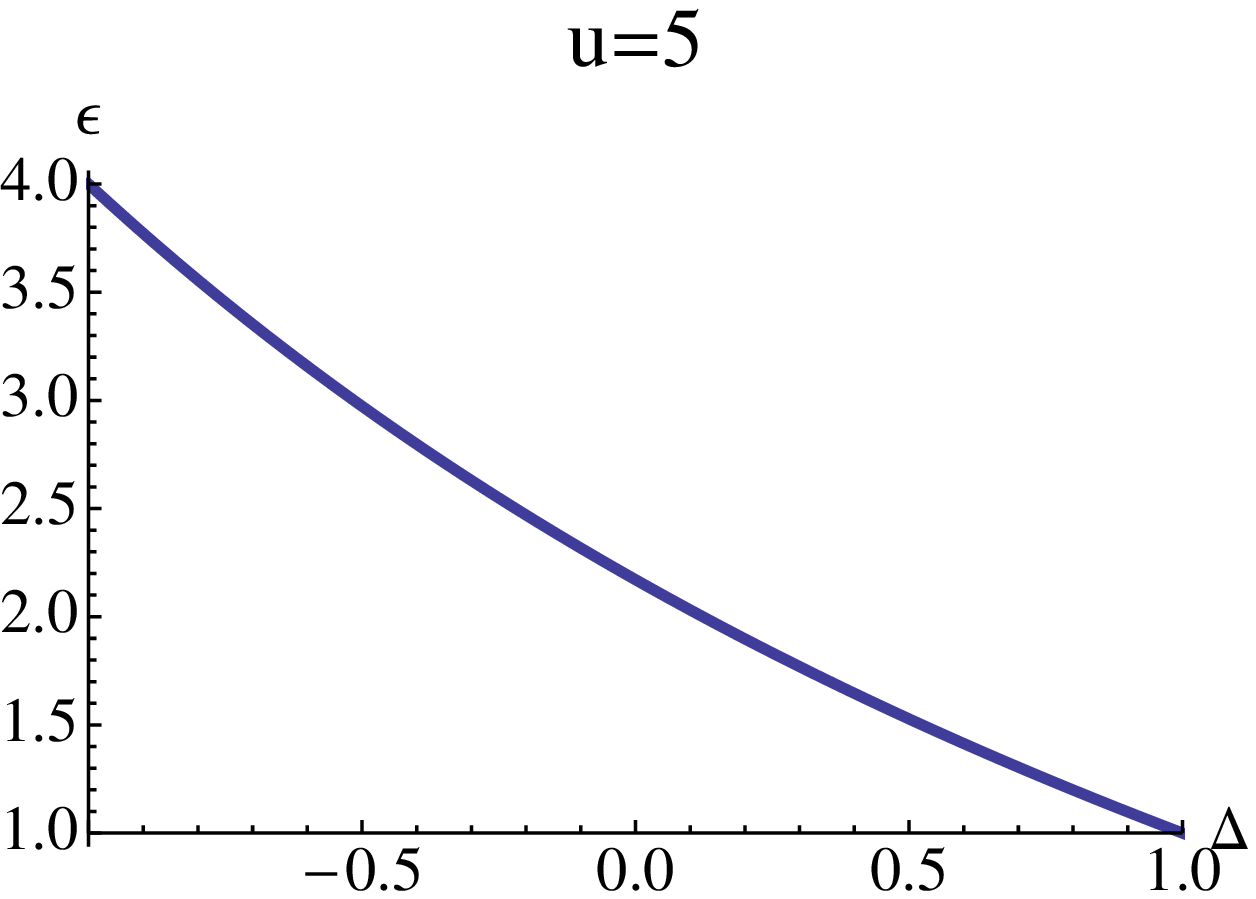}
  \hspace{2mm}
  \includegraphics[width=0.45\columnwidth]{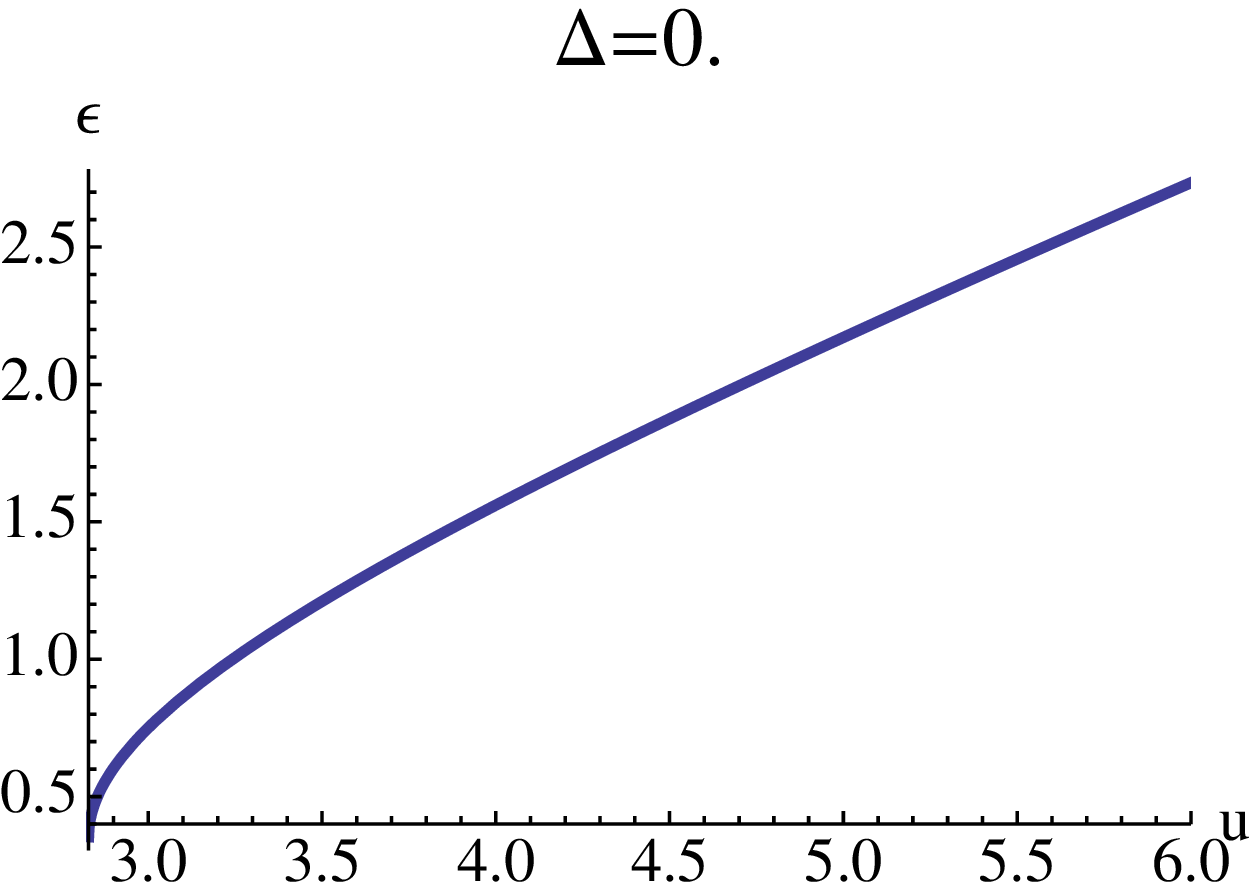}
\caption{The ground-state energy $\epsilon$ for $c=3$ is plotted against $\Delta$ for $u=5$ (left) and against $u$ for $\Delta=0$ (right).}
\Lfig{epsilon}
\end{center}
\end{figure}

The second moment $m_2$ shows more complicated behaviors. It diverges at $u=2\sqrt{c-1}$ and starts to decrease as $u$ grows from $2\sqrt{c-1}$, but for large $u$ it is a increasing function of $u$. Thus, there is an extremum for the region $u>2\sqrt{c-1}$. Similarly, for $u$ enough small but still larger than $2\sqrt{c-1}$, a non-monotonic behavior of $m_2$ with respect to $\Delta$ is observed. We plot those behaviors in \Rfig{m_2}.   
\begin{figure}[htbp]
\begin{center}
  \includegraphics[width=0.45\columnwidth]{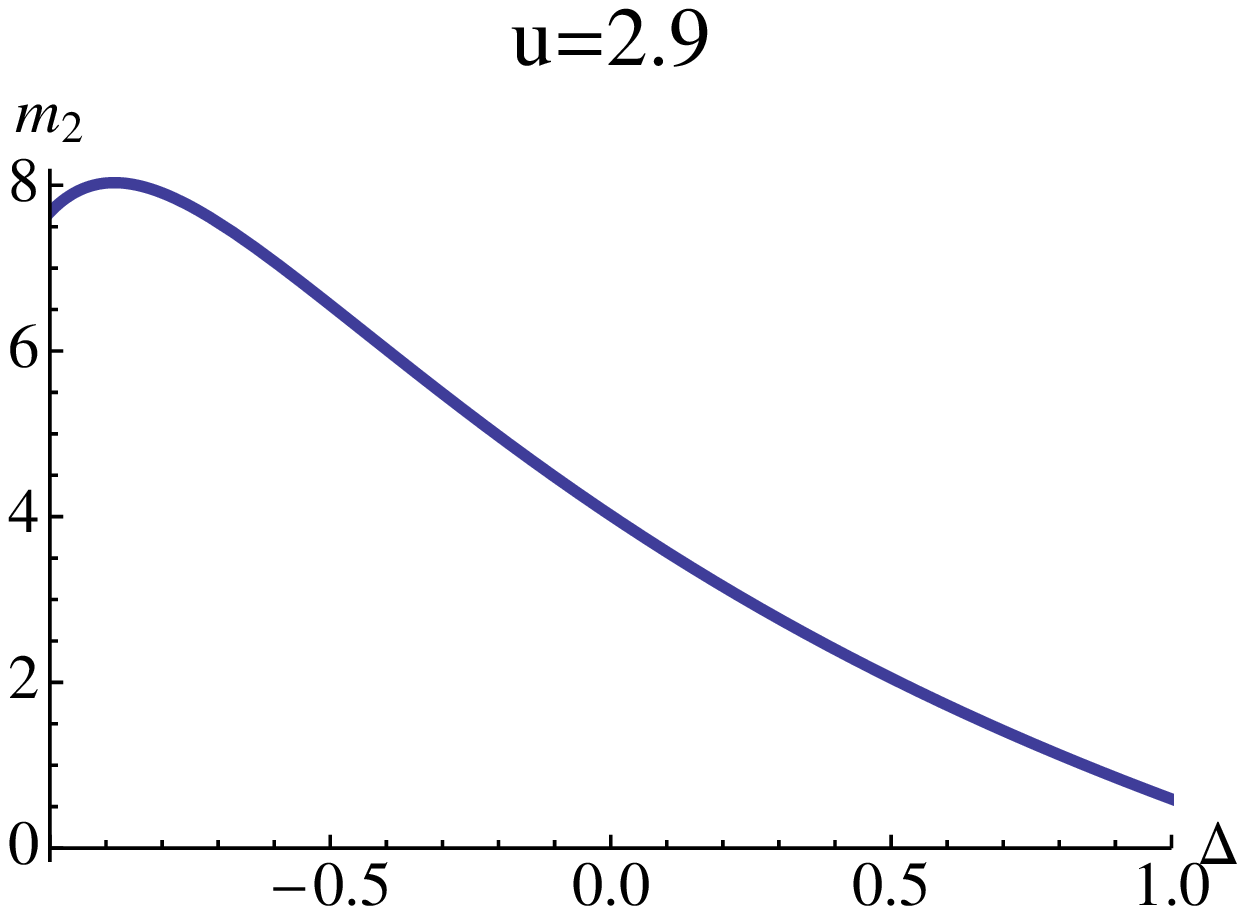}
  \hspace{2mm}
  \includegraphics[width=0.45\columnwidth]{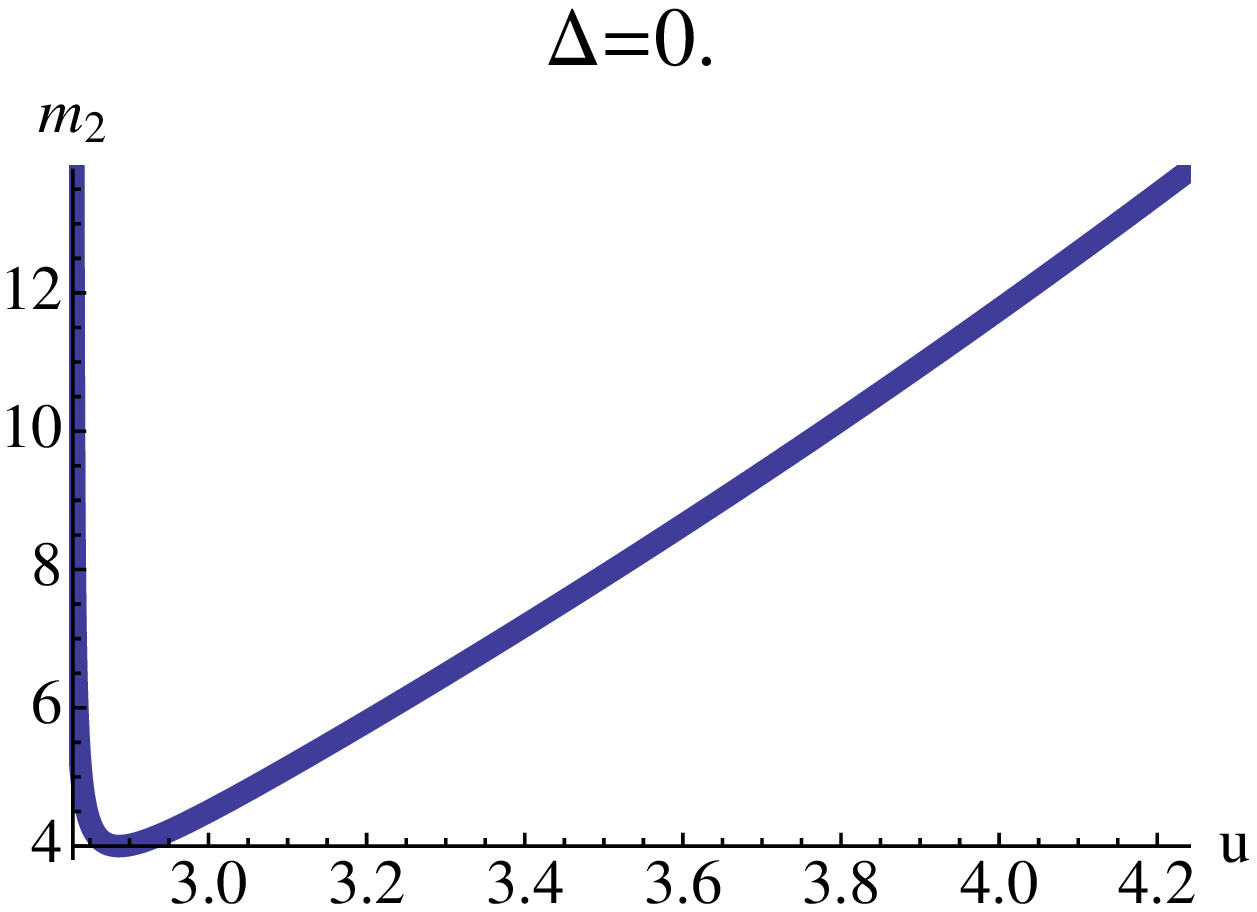}
\caption{The second moment $m_2$ for $c=3$ is plotted against $\Delta$ for $u=2.9$ (left) and against $u$ for $\Delta=0$ (right).}
\Lfig{m_2}
\end{center}
\end{figure}
On the other hand, these non-monotonic behaviors may not be meaningful for the ecological context, since there emerge extinct species for a certain value of $u$ sufficiently larger than $2\sqrt{c-1}$, and the present solution based on the Gaussian approximation does not give a precise result for such a small $u$. Hence, $m_2$ is basically considered to be a decreasing and increasing function with respect to $\Delta$ and $u$, respectively, in the meaningful region of $u$. 

There is another aspect worth noting. The above computations parallel that shown in the reference~\cite{Kabashima:12}, but there are two major differences: the self interaction is purely an external parameter (not the Lagrange multiplier to fix the vector norm) and there exists the Lagrange multiplier $r$ to hold $\sum_{i}x_i=N$ which breaks the rotational symmetry of $\V{x}$. Due to these differences, there does not occur a phase transition concerning the localization of $\V{x}$ occurring in the first eigenvalue problem~\cite{Kabashima:12}. An important consequence of this fact is the robustness of our result. Therefore, even if the degree of the network fluctuates, the result will be qualitatively the same as in the present case of the single degree of network $c$, in contrast to~\cite{Kabashima:12}. This robustness is numerically  observed and reported in~\cite{Obuchi:16}.

\subsection{The abundance distribution and related quantities}
In contrast to the order parameters calculated above, the abundance distribution requires the full functional forms of $q(H)$ and $\Wh{q}(\Wh{H})$. Based on the cavity interpretation stated above, the abundance distribution $P(x)$ is given by
\be
\hspace{-10mm}
P(x)=\int dA dH\, \delta\lb x- \frac{H}{A}\rb Q(A,H)
=
\int \prod_{l=1}^{c} d\Wh{H}_l \Wh{q}(\Wh{H}_l) \, \delta\lb x- \frac{r+\sum_{l=1}^{c}\Wh{H}_l}{u-c\hat{a}}\rb .
\ee
The functional form of $\Wh{q}(\Wh{H})$ is nontrivial. A standard way to obtain this is through a numerical technique called the population method. For simplicity of explanation, we here write down the self-consistent equation only of $\Wh{q}(\Wh{H})$ by using \Reqs{F2B}{B2F}:
\be
\Wh{q}(\Wh{H})=\int\prod_{l=1}^{c-1}d\Wh{H}_{l}\Wh{q}(\Wh{H}_{l})
\lsb 
\delta \lb \Wh{H}-\frac{J}{a} \lb r+\sum_{l=1}^{c-1}\Wh{H}_{l}  \rb   \rb 
\rsb_{J}.
\Leq{B2B}
\ee
In the population method, we parameterize the distribution $\Wh{q}(\Wh{H})$ by a large number of variables $\{\Wh{H}_{i}\}$, namely these variables should be distributed according to $\Wh{q}(\Wh{H})$. To achieve this, we recursively update the set of variables by the self-consistent equation \NReq{B2B}. The actual procedures are summarized as follows:
\begin{enumerate}
\item{Set an appropriate initial population of $\{\Wh{H}_{i}\}$ of size $N_{pop}$. We typically set $N_{pop}=40000$ and generate the population from the uniform distribution on $\lsb 0,1\rsb$.}
\item{Generate a new set of cavity biases each component of which is calculated from $c-1$ variables randomly chosen from the previous set of cavity biases with an interaction $J$ generated from \Req{J-dist}, according to the delta function in \Req{B2B}. The size of the new set is again $N_{pop}$. }
\item{Repeat (ii) until the distribution of the variables converges. The typical number of recursions we choose is $N_{rec}=40$. }
\end{enumerate}
This procedure constitutes a Markov chain of dynamics of the set of variables which is known to converge to the solution of the self-consistent equation. Using this convergent solution of $\Wh{q}(\Wh{H})$, we can evaluate the abundance distribution $P(x)$ and other related quantities. 

For sufficiently large $u$, the support of $P(x)$ is at $x>0$ and there are no extinct species. As $u$ decreases, the lower limit of the support becomes lower and lower, and at a certain value of $u_{}$ the support touches the point $x=0$. This defines the transition point $u_c$. Below this critical value $u \leq u_c$, there exist extinct species which are reflected in finite $P(x)$ in the negative $x$ region in the Gaussian approximation. Here we interpret $C(0)$, where we define the cumulative distribution as $C(y)=\int_{-\infty}^{y} dx P(x)$, as the proportion of the extinct species to the total population. According to this interpretation, we define the following modified distribution 
\be
\tilde{P}(x)=\theta(x)P(x)+C(0)\delta(x),
\ee
where $\theta(x)$ is the Heaviside step function. Unfortunately, the above interpretation is just an approximation for $u<u_c$, and the resultant abundance distribution $\tilde{P}(x)$ shows a deviation from the genuine distribution of the corresponding RD. Quantitative information on the deviation will be displayed later. 

We here enumerate other interesting quantities studied in this paper. The survival function is given by $\alpha(x)=1-C(x)$, which quantifies the proportion of species whose abundance is larger than $x$. Two special values of the survival function, $\alpha(0)$ and $\alpha(1)$, each of which corresponds to the proportions of surviving species and of species more abundant than the average, respectively, are used to measure the diversity of the community. The rank-abundance relation $x(r)$ is defined by the inverse function of the survival function as $x \lb r \rb =\alpha^{-1}(r)$. We display the actual behaviors of these quantities for several different $u$ and $\Delta$ below.

\subsubsection{Behaviors of the abundance-relating quantities}
In this section, we see the behaviors of the quantities explained above. The connectivity $c$ is fixed as $c=3$ since the qualitative behavior does not change by changing $c$. 

We start from the diversity $\alpha(0)$ and $\alpha(1)$, which are plotted against $\Delta$ and $u$ in \Rfigs{diversity-u}{diversity-delta}, respectively.   
\begin{figure}[htbp]
\begin{center}
  \includegraphics[width=0.32\columnwidth]{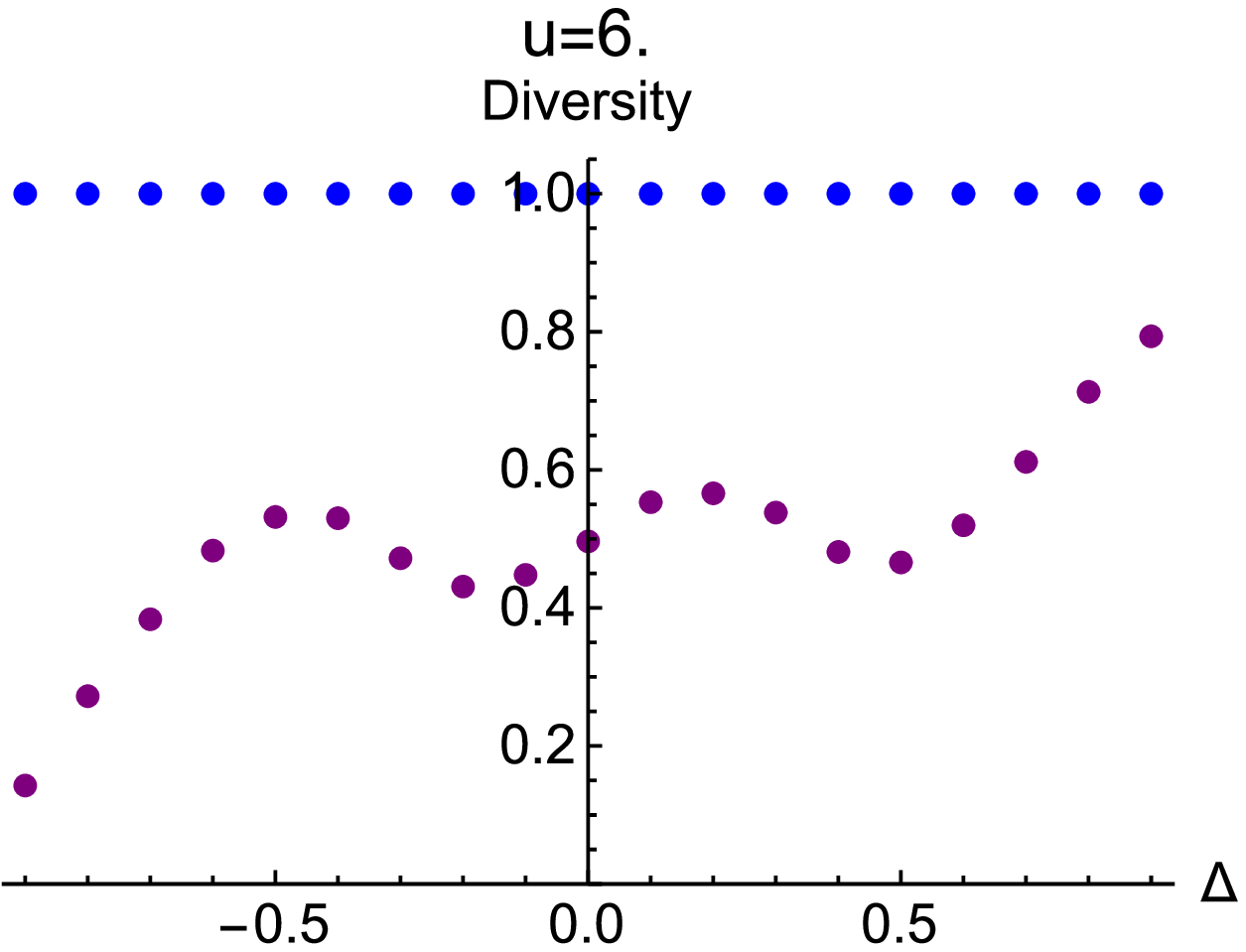}
  \includegraphics[width=0.32\columnwidth]{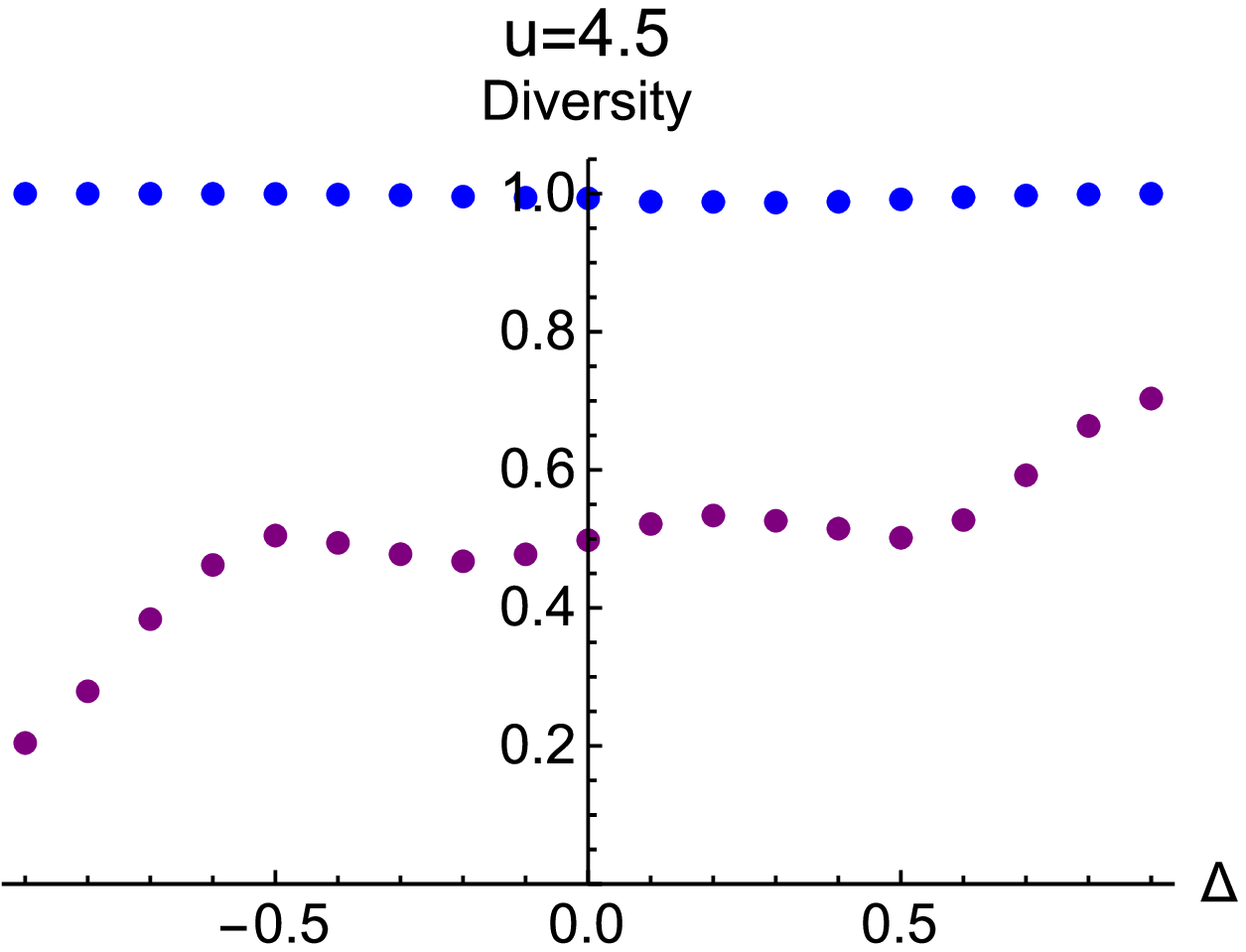}
  \includegraphics[width=0.32\columnwidth]{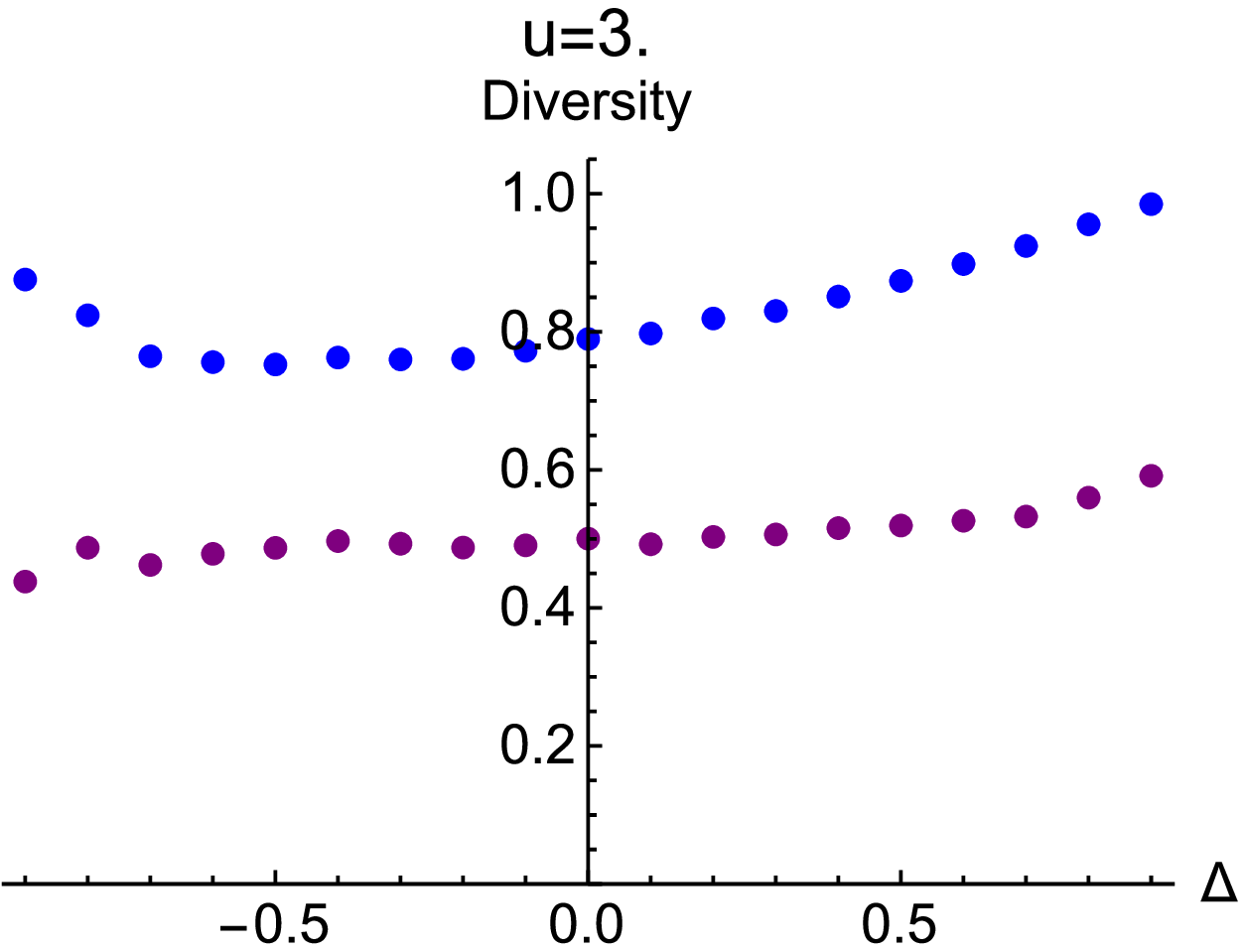}
\caption{The diversity $\alpha(0)$ (upper, blue) and $\alpha(1)$ (lower, purple) are plotted against the  $\Delta$ for different $u$. The dependence on $\Delta$ is not monotonic. }
\Lfig{diversity-u}
\end{center}
\end{figure}
\begin{figure}[htbp]
\begin{center}
  \includegraphics[width=0.32\columnwidth]{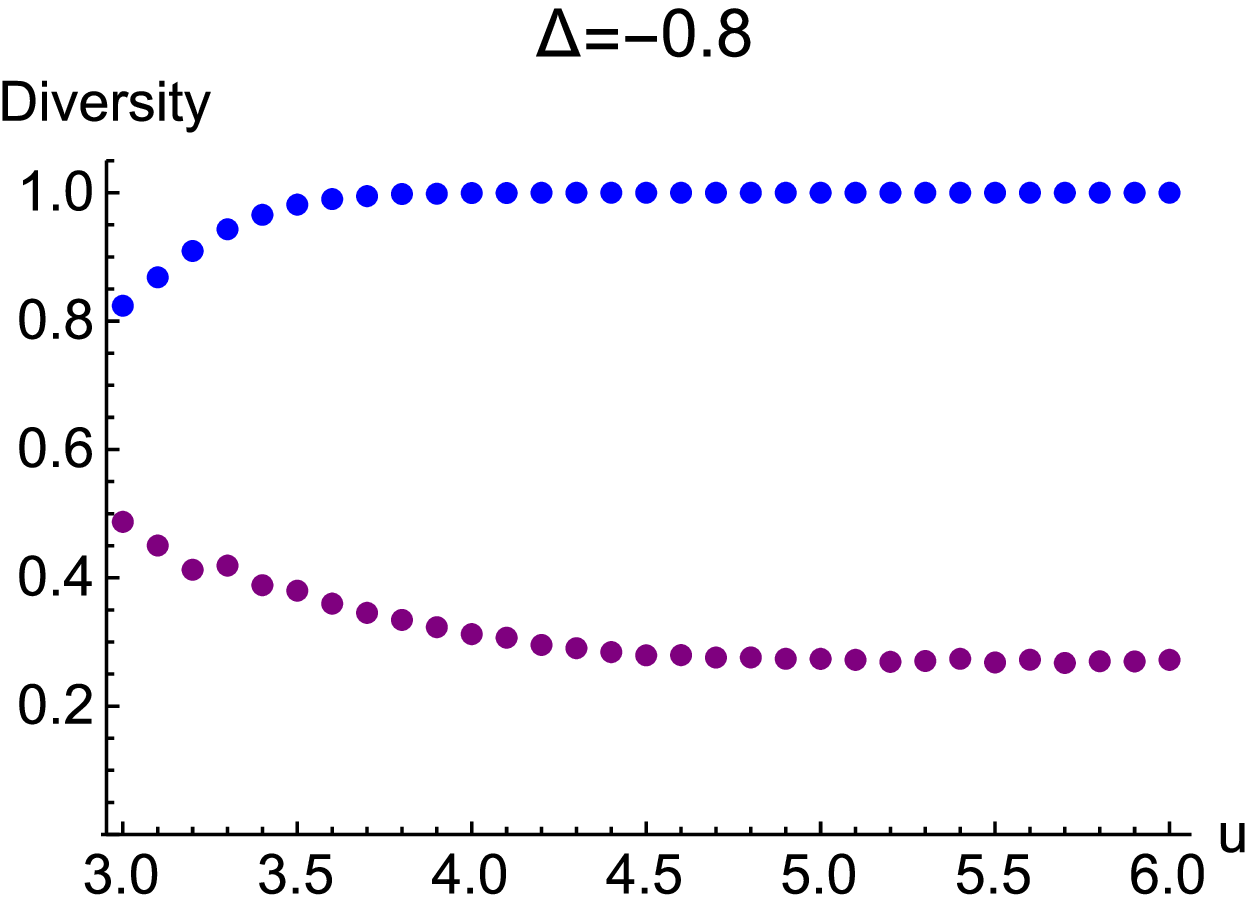}
  \includegraphics[width=0.32\columnwidth]{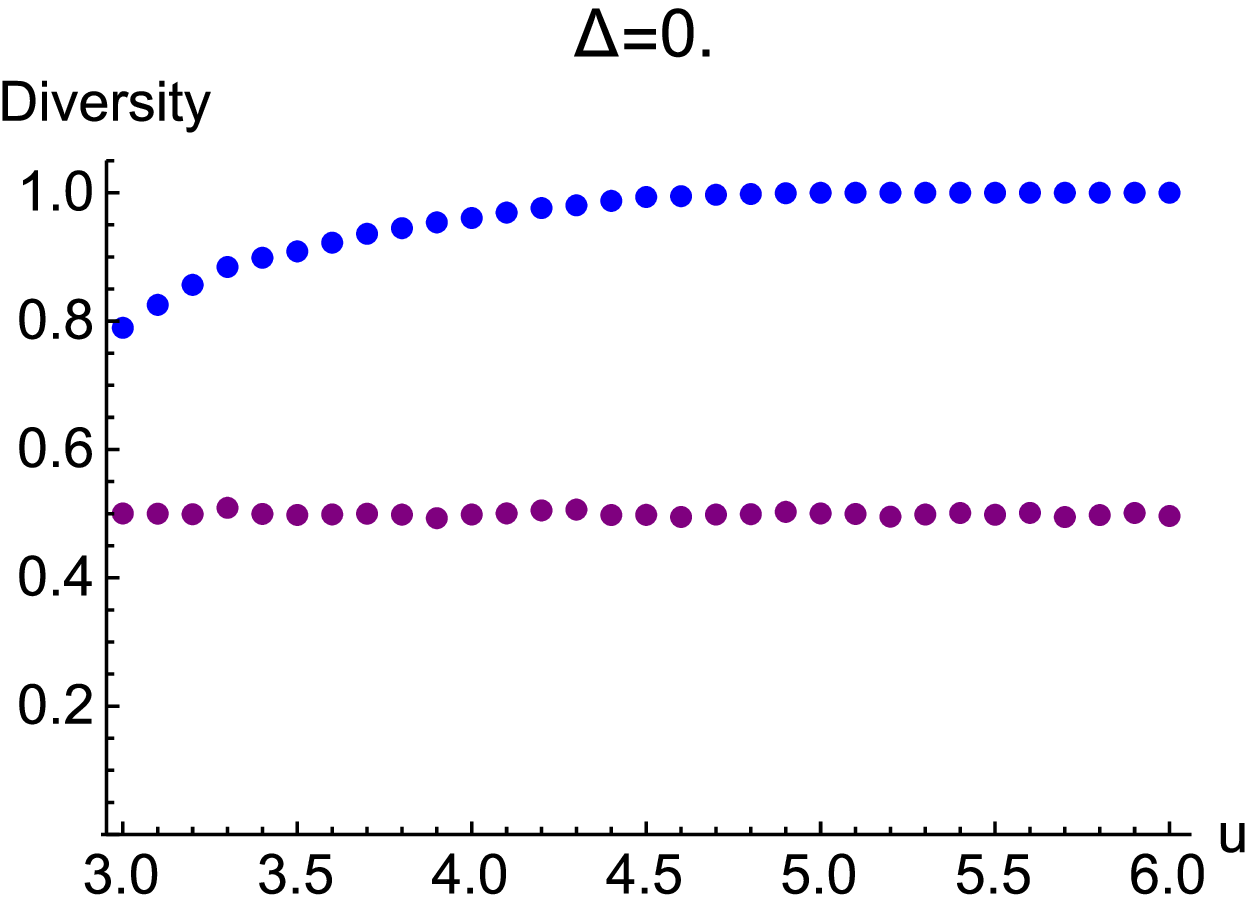}
  \includegraphics[width=0.32\columnwidth]{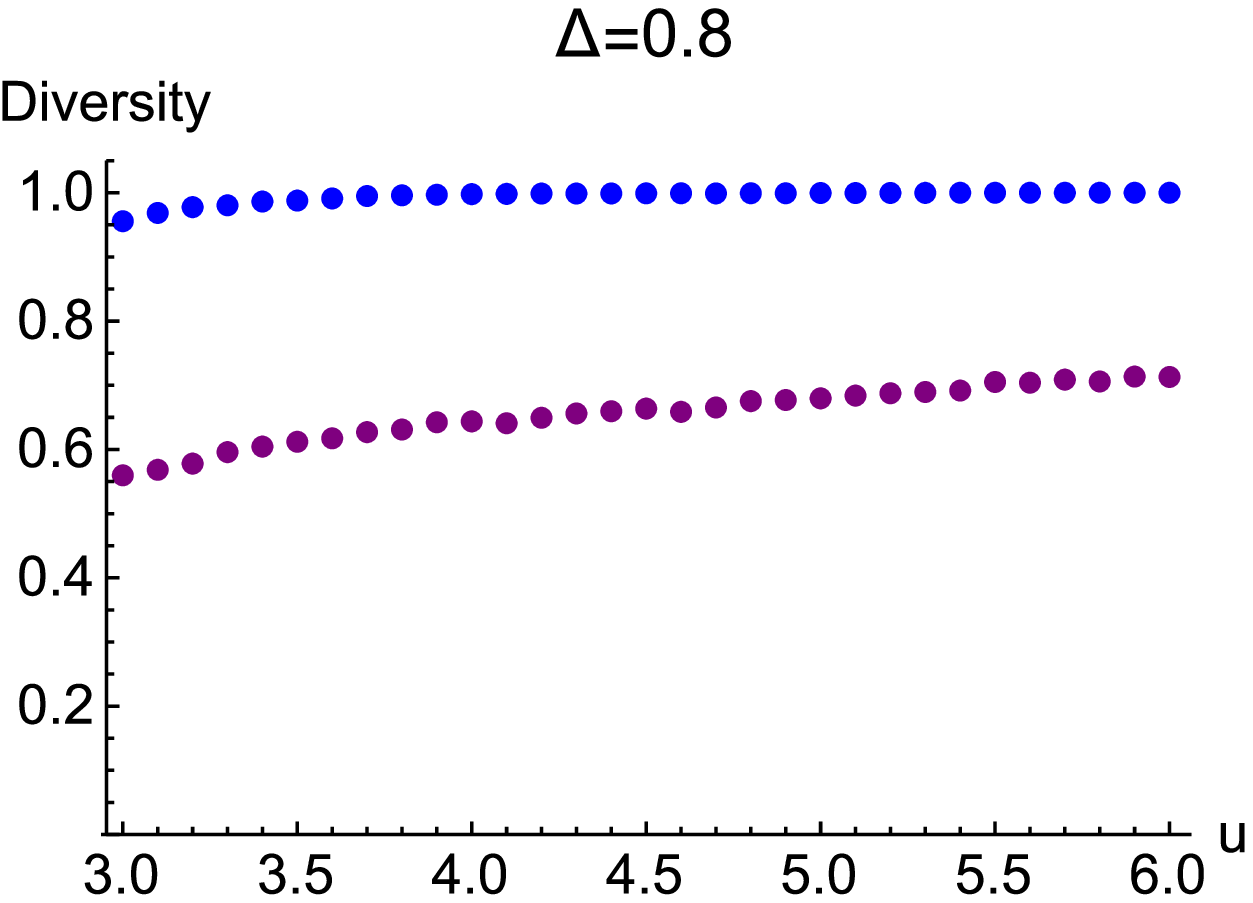}
\caption{The diversity $\alpha(0)$ (upper, blue) and $\alpha(1)$ (lower, purple) are plotted against the self interaction $u$ for different $\Delta$. }
\Lfig{diversity-delta}
\end{center}
\end{figure}
As seen in \Rfig{diversity-u}, the dependence of diversity on $\Delta$ is far from trivial. The oscillating behavior of $\alpha(1)$ is related to the multiple peaks of the abundance distribution appearing for large $u$. The height of each peak sensitively depends on $\Delta$; the location of the highest peak and the tail of the distribution changes as $\Delta$ deviates, which causes the oscillating behavior of $\alpha(1)$. Non-monotonicity of $\alpha(0)$ is interpreted as well. Meanwhile, an interesting observation from \Rfig{diversity-delta} is that the rich's diversity $\alpha(1)$ is a decreasing function of $u$ for the competitive case $\Delta=-0.8$ but is an increasing one for the mutualistic case $\Delta=0.8$, and is almost constant for the balanced case $\Delta=0$. In the context of evolution, this phenomenon implies that the mutualistic relation can motivate a boost in productivity $u$ in the community since many individuals can benefit from greater cooperation, though in a competitive community the opposite is the case. On the other hand, the survivor's diversity $\alpha(0)$ monotonically increases as $u$ grows and saturates to unity at the transition point $u=u_c$.

The transition point $u_c$ is plotted in the left panel of \Rfig{PD} against $\Delta$. To see the quantitative dependence of $u_c$ on the connectivity, we also plot $u_c$ against the connectivity $c$ for $\Delta=0$ in the right panel of the same figure.   
\begin{figure}[htbp]
\begin{center}
  \includegraphics[width=0.45\columnwidth]{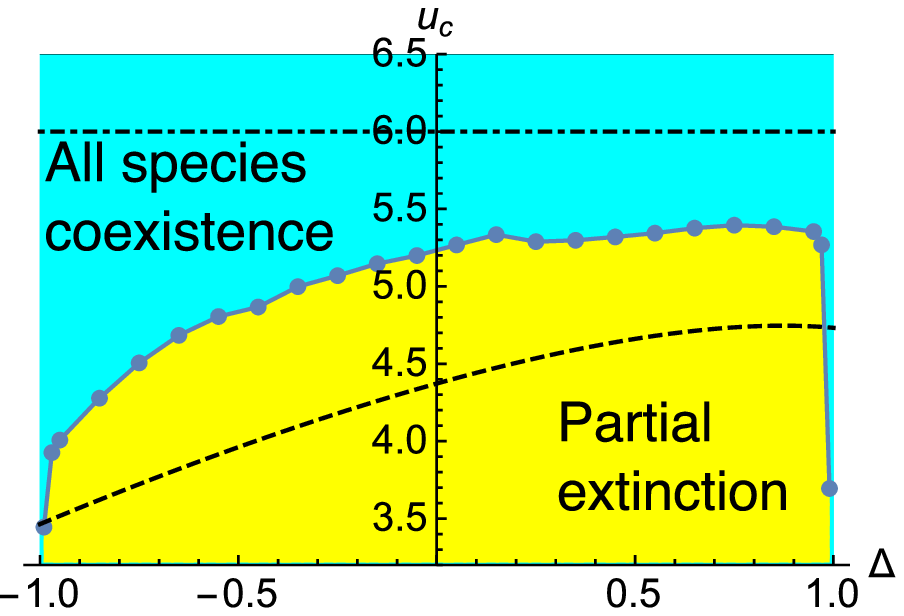}
  \includegraphics[width=0.45\columnwidth]{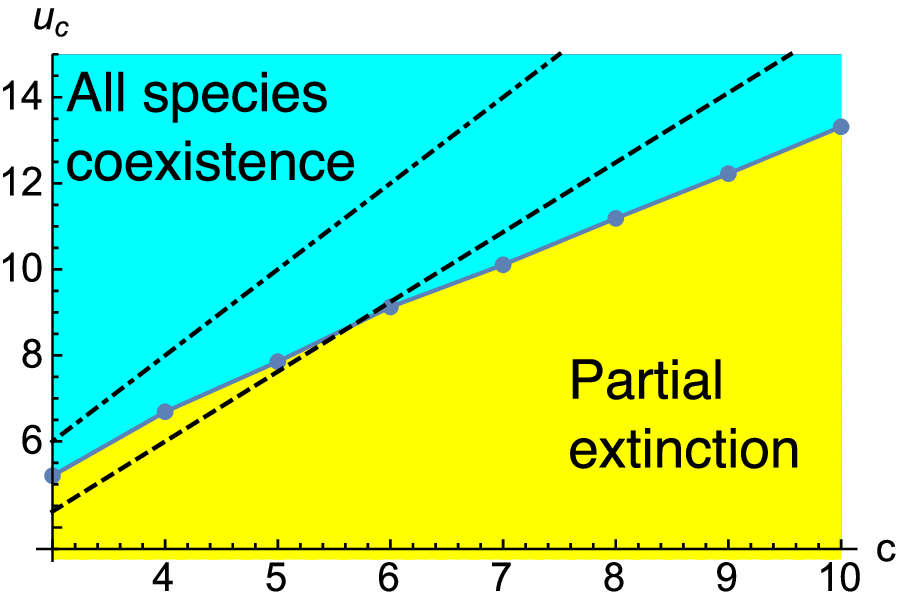}\caption{Transition points $u_c$ against $\Delta$ for $c=3$ (left) and against $c$ for $\Delta=0$ (right). The dots are the replica results, and the dashed lines correspond to the second-order perturbation approximation. Qualitative shapes of the $u_c$ curves are captured already by the second-order approximation. The dot-dashed lines represent the general upper bound $2c$ of $u_c$ derived in \Req{u_c-bound}, and both the perturbation and the replica results are certainly located below it. The critical value $u_c$ drastically drops off around $\Delta=\pm1$, which exhibits the singularity at those points.   }
\Lfig{PD}
\end{center}
\end{figure}
The approximation of $u_c$ by the expansion of $u^{-1}$ up to the second order is also plotted by the dashed lines in the figures. We can see the qualitative behavior is already captured by the second-order approximation, though the quantitative deviation is not small. The limits $\Delta \to \pm 1$ are singular: the value of $u_c$ drastically drops off around those limits as seen from the left panel. This behavior is expected: at $\Delta=\pm1$ all the species become equivalent since all the interactions take the same value, thus the abundance distribution should become $P(x)=\delta(x-1)$ from the symmetry, implying that $u_c$ has no meaning. Clear observation of this singularity is an advantage of the non-perturbative treatment since such a singularity is difficult to see with the perturbative expansion. From the right panel of \Rfig{PD}, we see the curve of $u_c$ is slightly jagged, which is seemingly due to numerical errors when solving \Req{B2B}. We have carefully examined the numerical accuracy with changing the parameters, and observed that this jagged behavior remains. Hence, we believe this jagged behavior actually occurs in the present model, which is presumably because of the non-monotonic dependence on $\Delta$ of the SAD.

Next, we examine the rank-abundance relations for several values of $u$ and $\Delta$. Here we choose $u=6, 4.5$ and $3$ since these three values locate above, close to, and below the transition point $u_c(\Delta)$, as seen from \Rfig{PD}. In the normal scale, the rank-abundance relations are given in \Rfig{abd-normal} for $\Delta=0.8,0$ and $0.8$.
\begin{figure}[htbp]
\begin{center}
  \includegraphics[width=0.32\columnwidth]{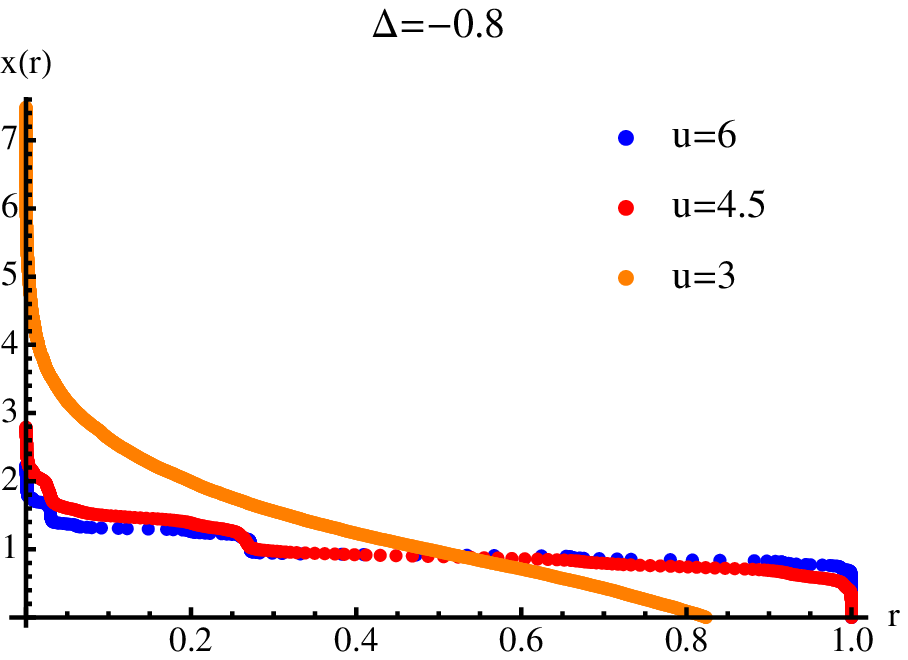}
  \includegraphics[width=0.32\columnwidth]{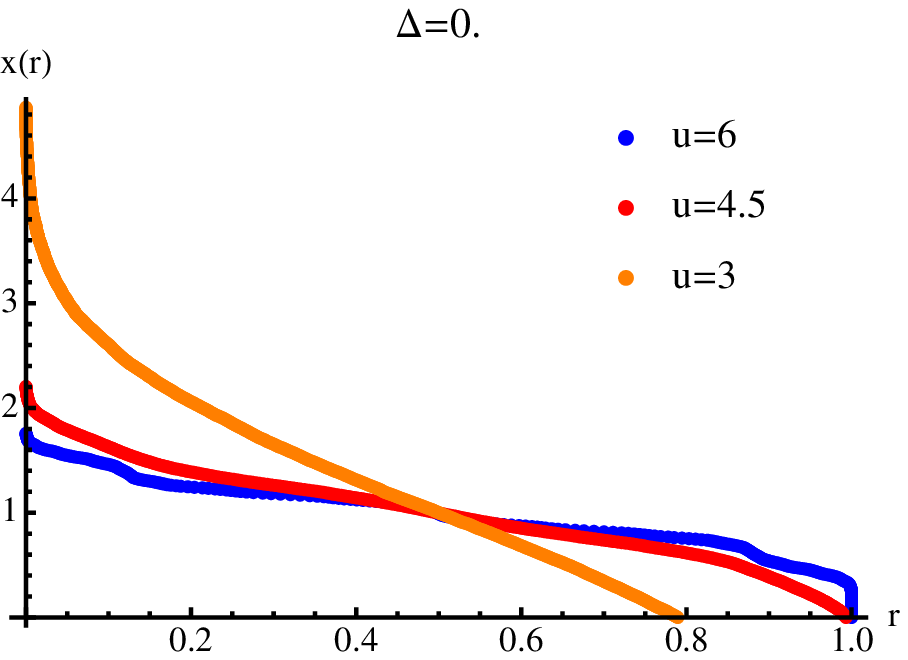}
  \includegraphics[width=0.32\columnwidth]{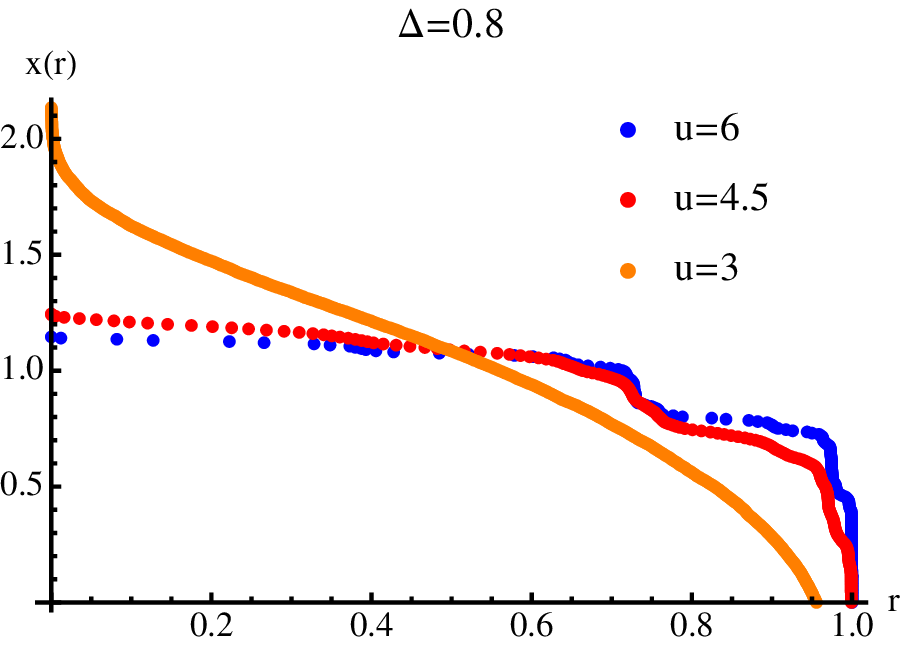}
\caption{The rank-abundance relations in the normal scale for $\Delta=-0.8,0$ and $0.8$ corresponding from left to right, for $u=3.0,4.5$ and $6$.}
\Lfig{abd-normal}
\end{center}
\end{figure}
By the discreteness of the abundance distribution stated in \Rsec{perturbation}, the rank-abundance relations show step-function-like behaviors for large $u$, but they are gradually rounded as $u$ decreases and the functional forms become like sigmoid functions (see $u=3$). This becomes clearer on a semi-logarithmic scale, which is given in \Rfig{abd-log}.
\begin{figure}[htbp]
\begin{center}
  \includegraphics[width=0.32\columnwidth]{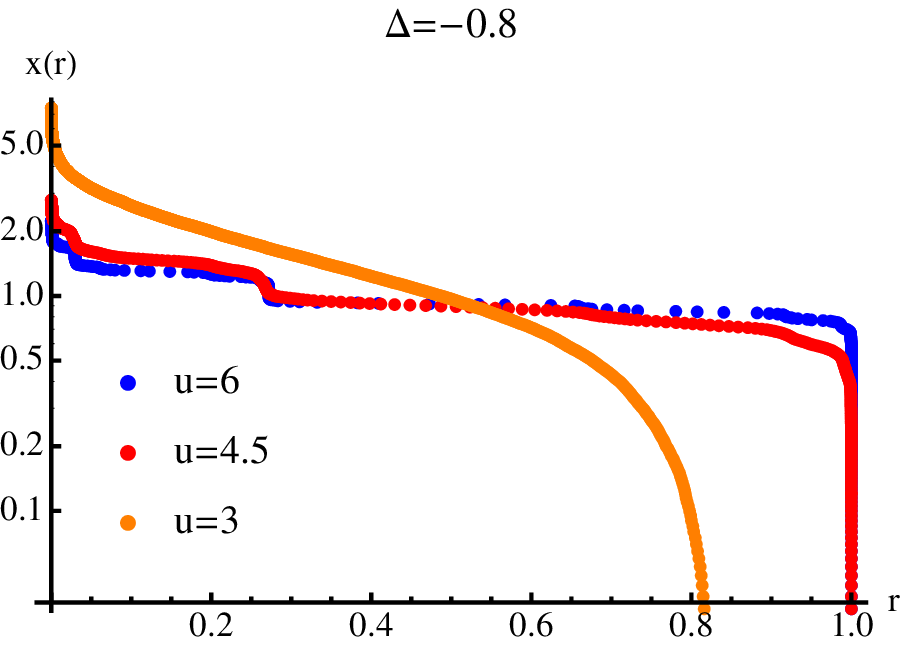}
  \includegraphics[width=0.32\columnwidth]{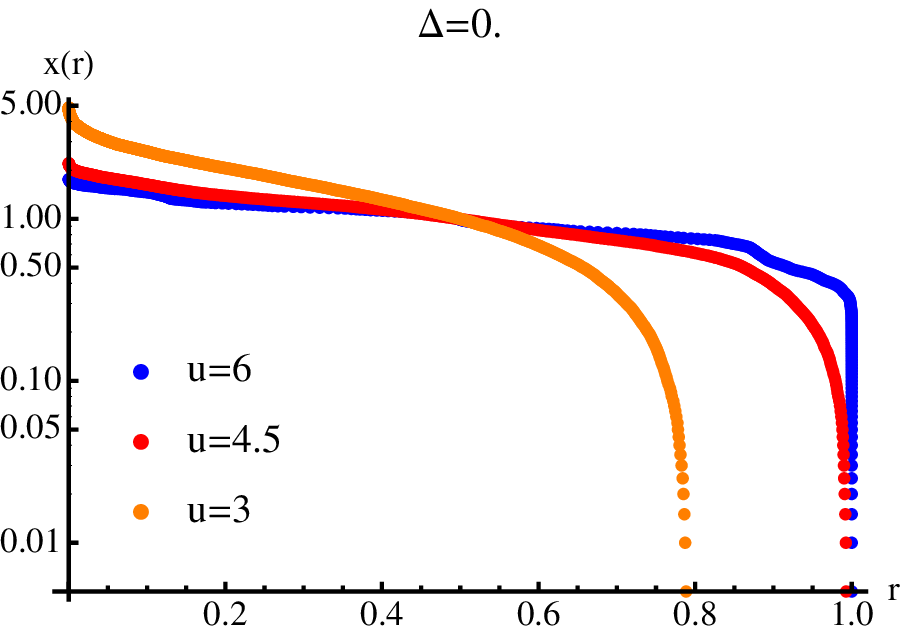}
  \includegraphics[width=0.32\columnwidth]{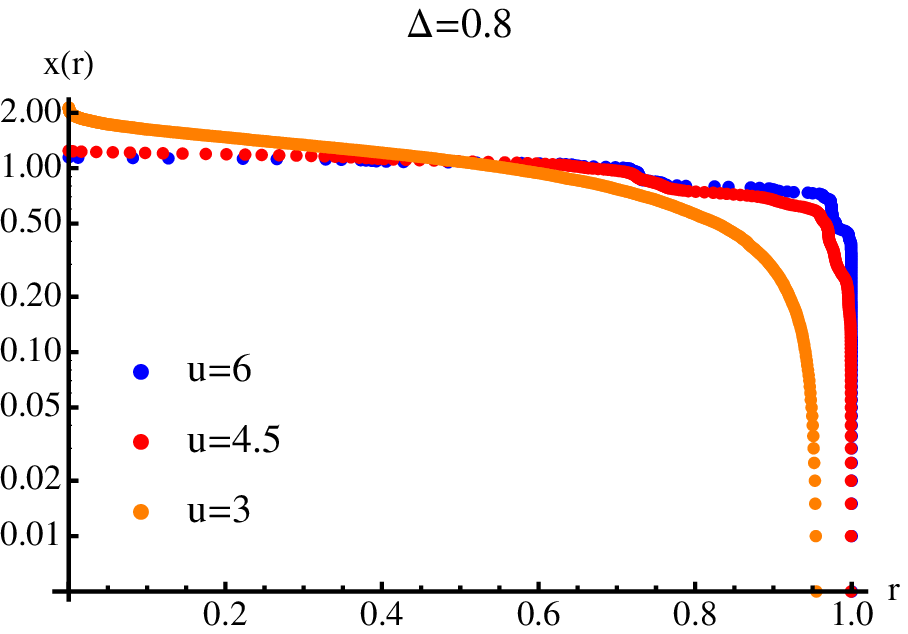}
\caption{The rank-abundance relations in the semi-logarithmic scale for $\Delta=-0.8,0$ and $0.8$ corresponding from left to right, for $u=3.0,4.5$ and $6$.}
\Lfig{abd-log}
\end{center}
\end{figure}

The abundance distributions are summarized in \Rfig{hist-multi}. We see clear discreteness in the abundance distributions for large $u$. Though this is suggested already by the perturbative analysis, the quantitative information free from the perturbative approximation is useful. We point out that the value of $u=6$ is equal to the general upper bound of $2c$ derived from \Req{u_c-bound}, and thus not so large. Hence, we can conclude that the discreteness of the abundance distribution survives well even for reasonable values of the productivity $u$. 
\begin{figure}[htbp]
\begin{center}
  \includegraphics[width=0.32\columnwidth]{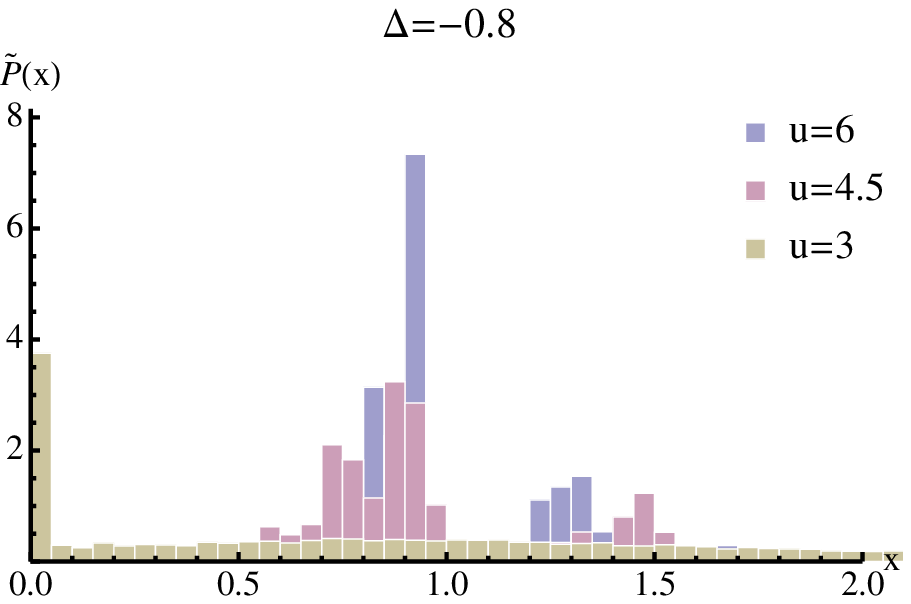}
  \includegraphics[width=0.32\columnwidth]{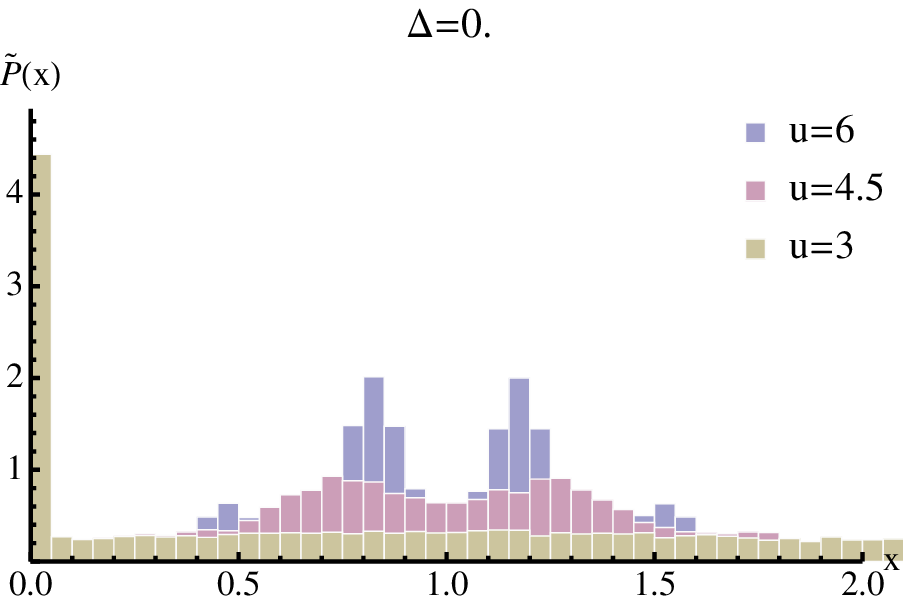}
  \includegraphics[width=0.32\columnwidth]{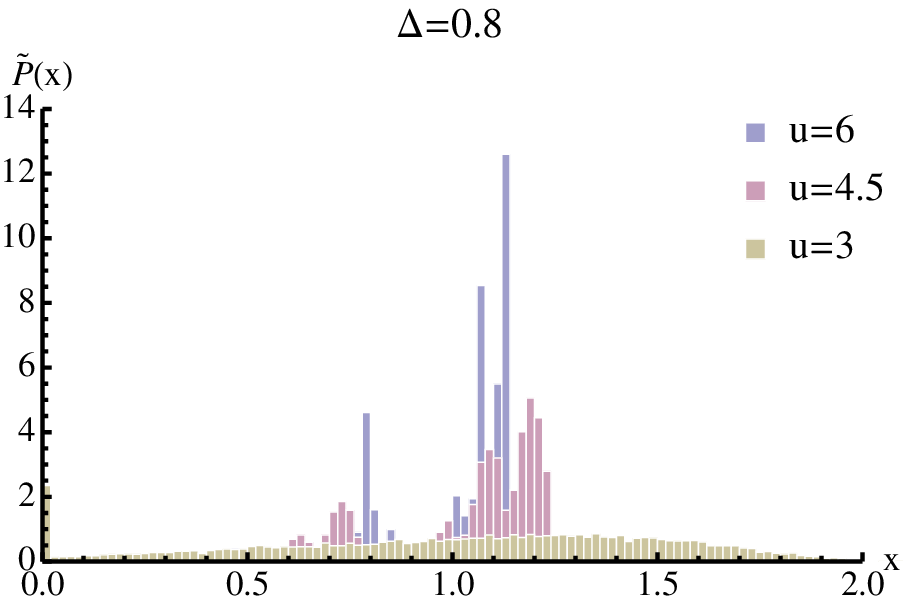}
\caption{The abundance distributions for $\Delta=-0.8,0$ and $0.8$ corresponding from left to right.}
\Lfig{hist-multi}
\end{center}
\end{figure}
For $u=6$, the distribution is symmetric about $x=1$ for $\Delta=0$, though the distribution is biased to $x>1$ or to $x<1$ for $\Delta \neq 0$. For the competitive case $\Delta=-0.8$, the largest peak appears in $x<1$, and the long tail persists in the $x>1$ region, while for the mutualistic case $\Delta=0.8$ the opposite is true. These results accord with the perturbation predictions. As $u$ decreases, the discreteness becomes weaker and the extinct species starts to emerge, and the functional forms gradually tend to become similar among different $\Delta$. 

\subsubsection{Comparison with direct simulations of the RD}
We work on the Gaussian approximation which will give incorrect results in $u<u_{c}$. To observe this deviation from the correct result, we also perform the direct simulation of the RD on RRG of $c=4$ for large and small $u$. The results are given in \Rfigs{hist-comp}{abd-comp}. For the simulation, we numerically solve the RD on the RRG by the Runge-Kutta method of the fourth order. The initial condition is chosen as the uniform one $x_{i}=1,(\forall{i})$, and the system size is $N=16,000$; the finite-size effect on the abundance distribution is confirmed to be absent for $N\geq 8000$. The sample average is not taken since the fluctuation of the abundance distribution is small enough for this system size. 
\begin{figure}[htbp]
\begin{center}
  \includegraphics[width=0.45\columnwidth]{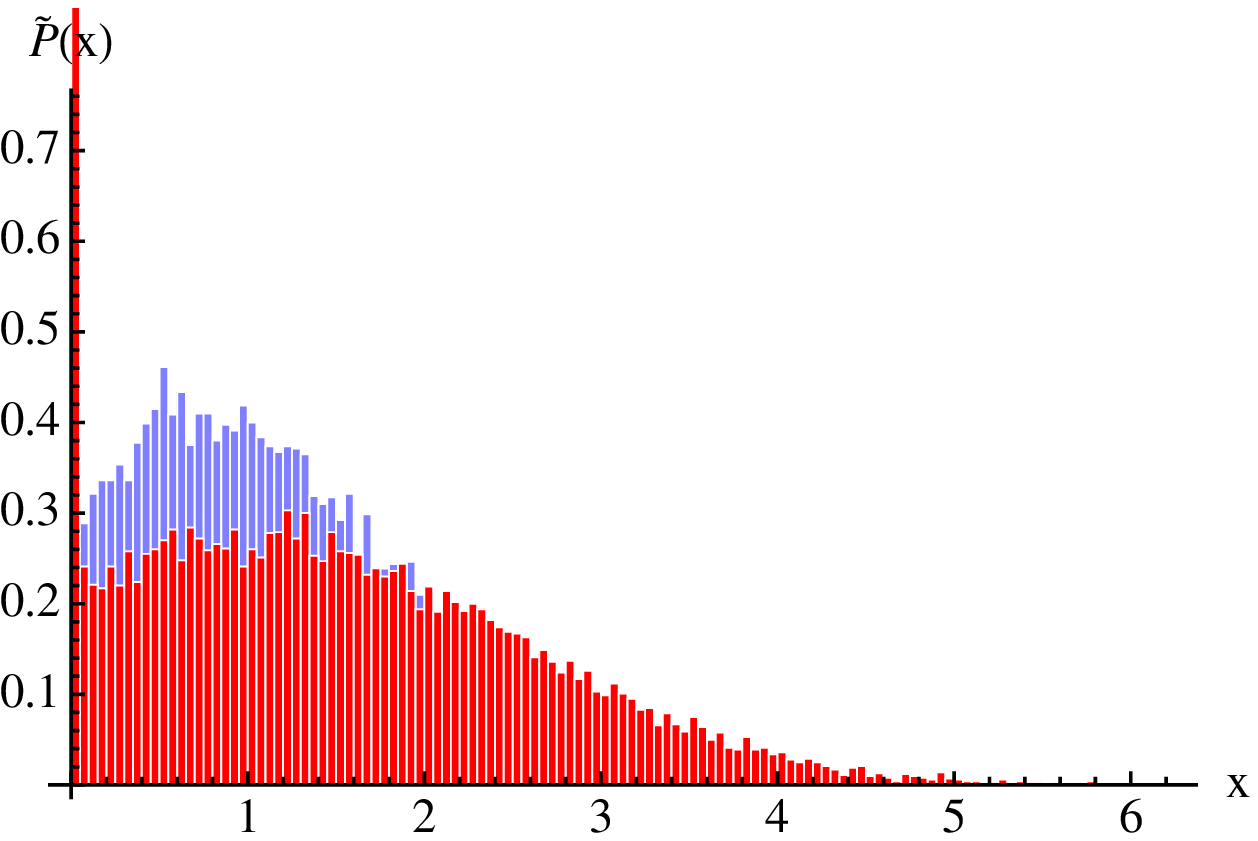}
  \includegraphics[width=0.45\columnwidth]{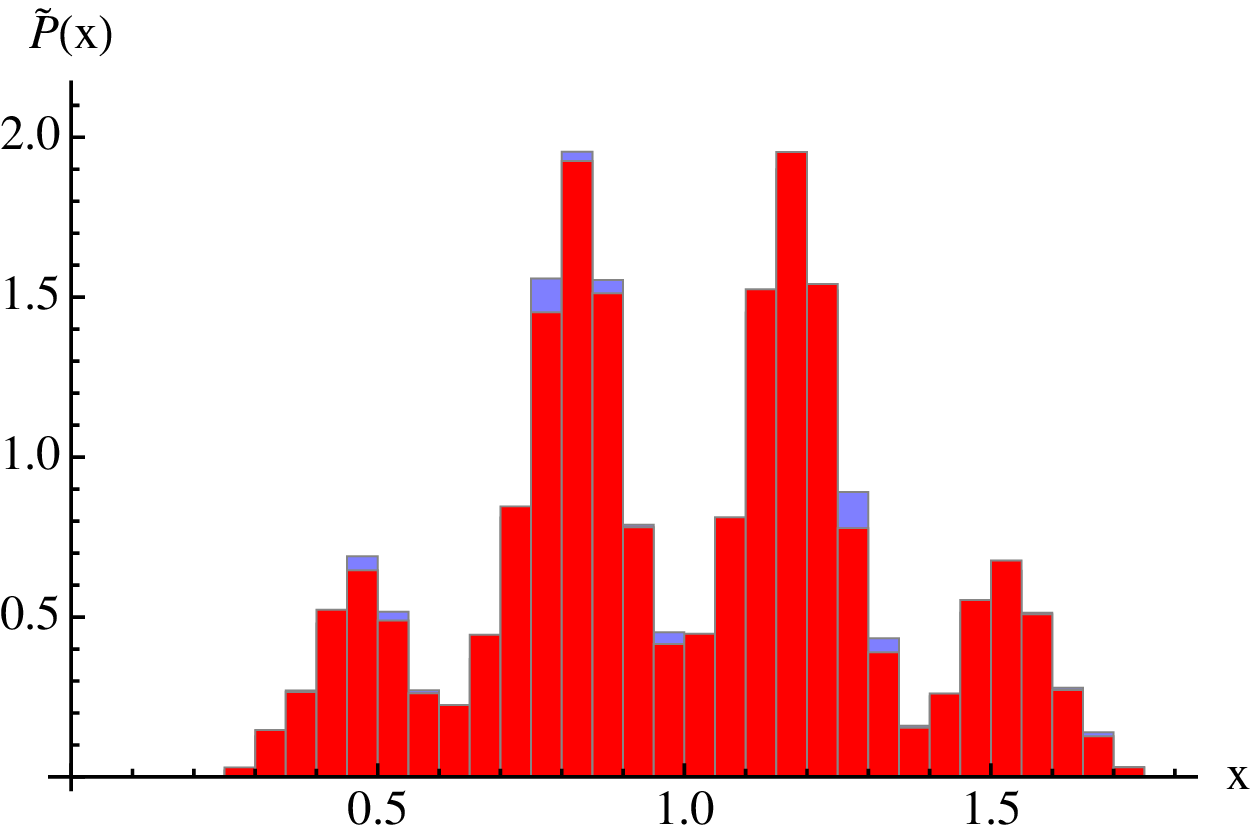}
\caption{The abundance distributions for $u=2.93$ (left) and $6.03$ (right) for $\Delta=0$. Red bars correspond to the replica result and the blue ones to that of the simulation. They accord for large $u$, but show some deviation for small $u$ where extinct species exist.}
\Lfig{hist-comp}
\end{center}
\end{figure}
\begin{figure}[htbp]
\begin{center}
  \includegraphics[width=0.45\columnwidth]{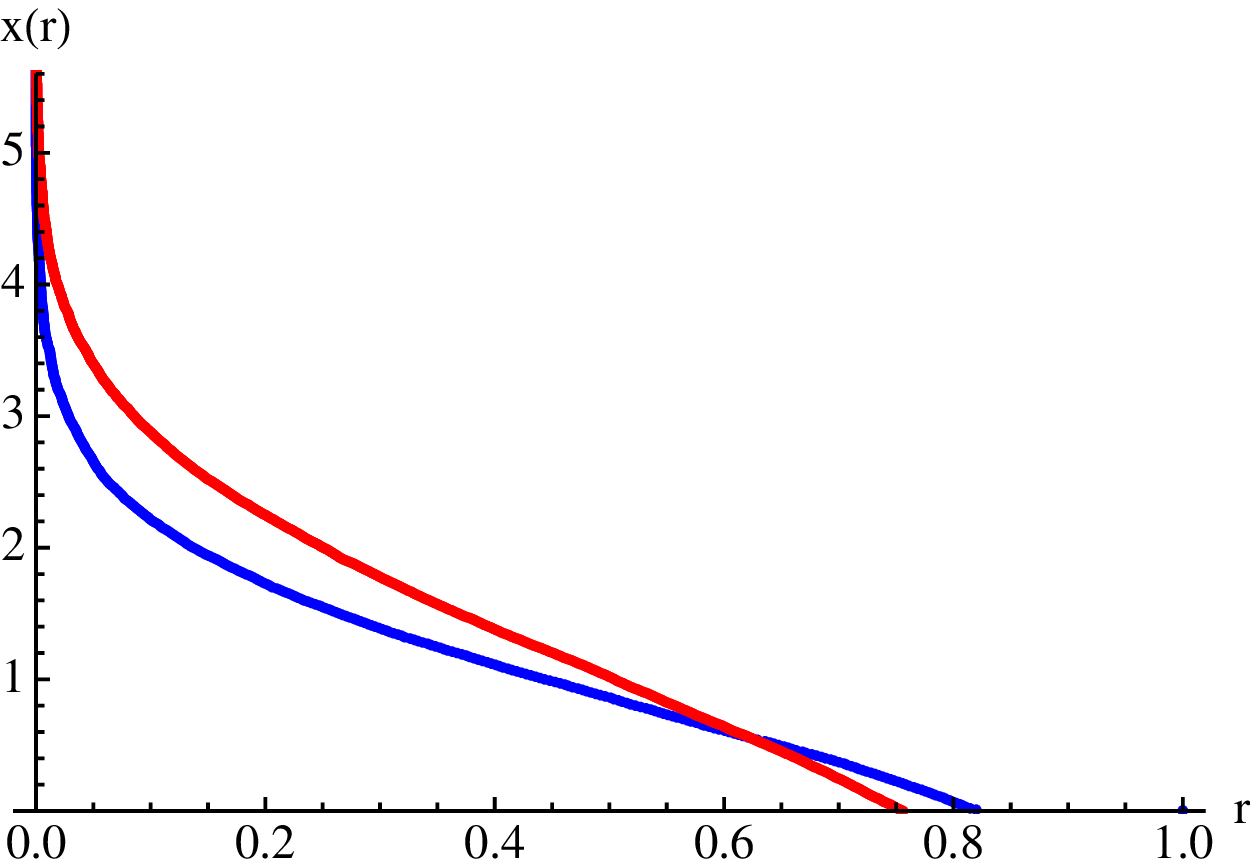}
  \includegraphics[width=0.45\columnwidth]{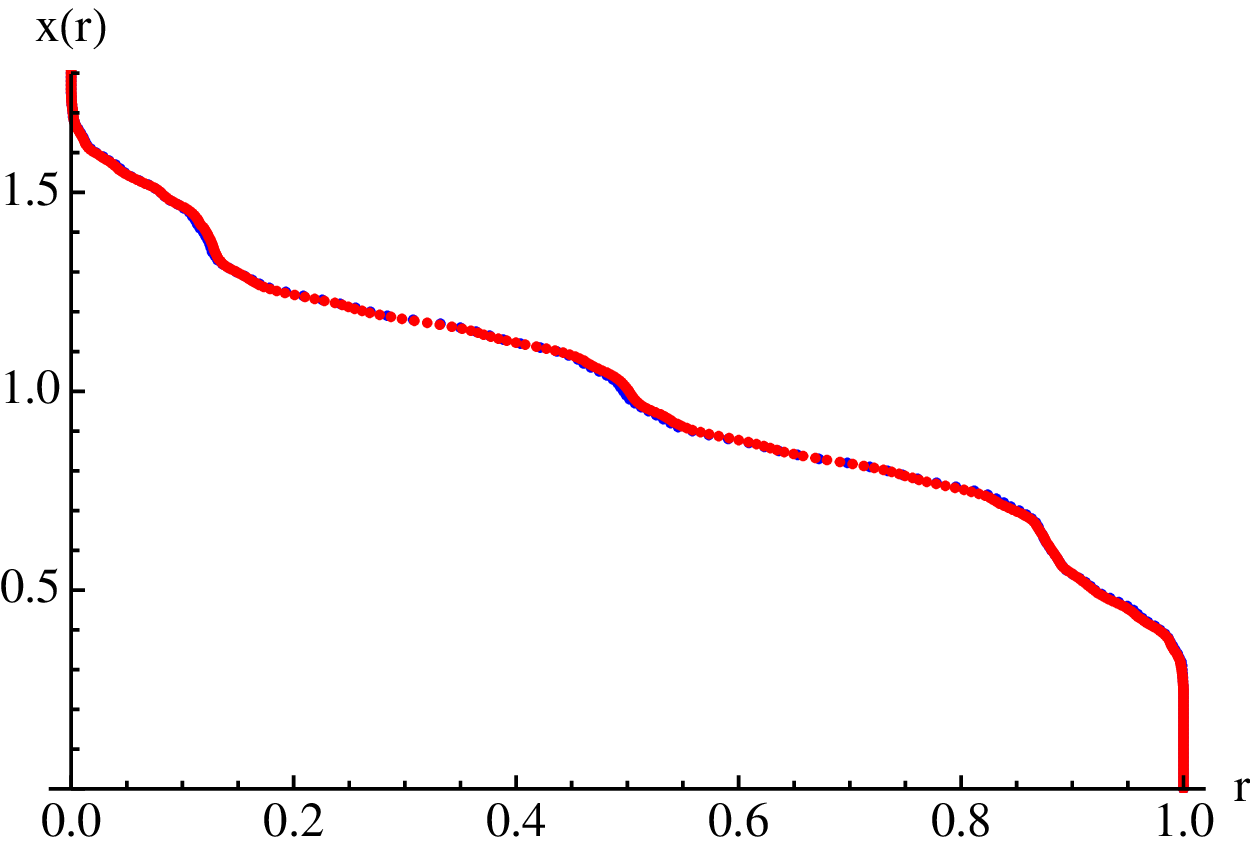}
\caption{Rank-abundance relations for different $u=2.93$ (left) and $6.03$ (right) for  $\Delta=0$. Red plots correspond to the replica result and the blue ones to that of the simulation.}
\Lfig{abd-comp}
\end{center}
\end{figure}
We see the complete accordance for large $u=6.03$ in the right panels of \Rfigs{hist-comp}{abd-comp}, though there is a deviation in the left panels of those figures of small $u=2.93$. 

\section{Conclusion}\Lsec{Conclusion}
In this paper, we investigated the RD on sparsely connected networks with symmetric interactions by studying the global minimum of the Hamiltonian corresponding to the Lyapunov function, the existence of which  guarantees the convergence of the RD dynamics to fixed points. The sparseness of the interaction network produces a wide variety of RSA patterns, though the analytical treatment becomes more difficult since the self-consistent relation to derive the single-site marginal distribution becomes complicated. We did not tackle this problem directly and instead treated the model with large self interactions, which enabled us to treat it in a very systematic manner. Thanks to the large self interactions, we could neglect the non-negativity constraint of the population $x_{i}\geq 0$, and the direct minimization of the Hamiltonian was possible. The resultant formula is appealingly simple and we constructed a perturbative expansion with respect to the inverse of the self interaction. This was reformulated in terms of the Boltzmann distribution with the aid of the Gaussian approximation. Working on this approximation, we invented a non-perturbative theory on the regular random graph and derived some characteristic RSA patterns. This non-perturbative treatment formally also works in the region of small $u$ region, though our result becomes not precise due to the failure of the Gaussian approximation by the presence of extinct species, as clarified by the comparison with the direct simulations of the RD. 

The derived RSA patterns directly reflect the nature of the interactions, in contrast to the fully-connected case. As an example, we treated binary interactions $J_{ij}=\pm 1$, which leads to multiple peaks in the abundance distribution for large $u$. Such multiple peaks were actually observed in some earlier experimental works, and it will be a promising future investigation to clarify the relation between the presented theory and those experimental data. For comparing with experimental works, the robustness of the multiple peaks by the present mechanism, the discreteness of the interactions on sparsely connected networks, is an crucial issue. We have conducted some additional numerical simulations and confirmed that they are fairly robust against a certain level of modification of the model parameters and the network structure, the result of which has been reported in~\cite{Obuchi:16}. This reinforces the plausibility of the presented mechanism of multiple peaks in the abundance distribution. 

Another interesting issue is the origin of the discreteness of the interactions. Although we do not have reasonable biological explanations or observations supporting the discreteness of the interactions, we can imagine that this possibly occurs if some species compete for a common resource, since the interactions among those competing species will be determined only by the resource. In any case, further investigation is desired on this issue. 

The multiple-peak distribution tends to become rounded as the self interaction decreases, and the resultant distribution's shape becomes close to that observed in the fully-connected case, which is clear in the sigmoid-type shape in the rank-abundance relation. We stress that this drastic change of the distribution's shape is controlled by a few parameters, the self interaction $u$ and the ratio of mutualistic relations $\Delta$. Thus our theory provides a possibility of unifying different shapes of the abundance distributions. This flexibility will help us to understand a wide variety of RSA patterns actually observed in many field research. 

\section*{Acknowledgments}
This work was supported by Grant-in-Aid for JSPS Fellows (No. 2011) (TO),  KAKENHI No. 26870185 (TO), 25120013 (YK), and 24570099 (KT). KT also acknowledges support in part by the project ``Creation and Sustainable Governance of New Commons through Formation of Integrated Local Environmental Knowledge'', at the Research Institute for Humanity and Nature (RIHN), and the project ``General Communication Studies'', at the International Institute for Advanced Studies (IIAS). 

\appendix
\section{Replica calculations on random graphs}
\Lsec{app:replica}

\subsection{The number of realizations of RRG with fixed connectivity $c$}
It is a good exercise to calculate the number of RRGs with fixed connectivity $\mathcal{G}$:  
\be
\hspace{-10mm}
\mathcal{G}=\sum_{ \{L_{ \Ave{ij} }=0,1 \} }
\prod_{k=1}^{N}\delta\lb \sum_{j\neq k}L_{\Ave{kj}}-c \rb
=
\sum_{ \{L_{ \Ave{ij} }=0,1 \} }
\prod_{k=1}^{N}\oint\frac{  dz_{k}  z_k^{  -(c+1)  }  }{  2\pi i  }
\prod_{i=1}^{N}z_{i}^{\sum_{j\neq i} L_{\Ave{ij}}},
\ee
where we used the identity
\be
\delta(x)=\oint\frac{dz }{2\pi i}z^{-(x+1)}.
\ee
The integration path is a closed one, enclosing the origin of the complex plane of $z$. The variable $L_{ \Ave{ij} }$ denotes the presence $(L_{ \Ave{ij} }=1)$ and absence $(L_{ \Ave{ij} }=0)$ of the link $\Ave{ij}$ on the graph. Here we perform the following transformation
\be
\hspace{-20mm}
\sum_{ \{L_{ \Ave{ij} }=0,1 \} }\prod_{i=1}^{N}z_{i}^{\sum_{j\neq i} L_{\Ave{ij}}}
=
\prod_{\Ave{ij}}\sum_{ L_{ \Ave{ij} }=0,1  }(z_{i}z_{j})^{L_{\Ave{ij}}}
=\prod_{\Ave{ij}}(1+z_iz_j)\approx \prod_{\Ave{ij}}e^{z_iz_j}\approx e^{\frac{1}{2}(\sum_{i}z_{i})^2}.
\Leq{graph-trick}
\ee
The Hubbard-Stratonovich transformation gives
\be
e^{\frac{1}{2}(\sum_{i}z_{i})^2}=A\int dx e^{-\frac{N}{2}x^2+\sqrt{N}x\sum_{i}z_i}.
\ee
The constant $A$ is irrelevant and will be discarded hereafter. For the integration with respect to $z_i$, the surviving term is only
\be
\oint \frac{dz z^{-(c+1)}}{2\pi i}e^{qz} 
= \oint \frac{dz z^{-(c+1)}}{2\pi i} \frac{q^{c}}{c!}z^{c}=\frac{q^{c}}{c!}.
\ee
Thus,
\be
&&\mathcal{G}=
\prod_{k=1}^{N}\oint\frac{  dz_{k}  z_k^{  -(c+1)  }  }{  2\pi i  }
e^{\frac{1}{2}\lb \sum_{i}z_{i} \rb^2}
=
\int dx 
e^{-\frac{N}{2}x^2}
\lbb 
\lb \frac{x^{c}}{c!}\rb^{N} 
\rbb
N^{Nc/2}
\no \\
&&
=\int dx \exp N\lbb 
-\frac{1}{2}x^2+\log\frac{x^{c}}{c!}+\frac{c}{2}\log N
\rbb.
\ee
The saddle-point condition gives $x^2=c$, and 
\be
\frac{1}{N}\log \mathcal{G}=-\frac{1}{2}c+\frac{1}{2}c\log cN -\log c!.
\ee

\subsection{Computation of the free energy}
We can perform similar transformations to the previous subsection for calculating the free energy
\be
&&[Z^n]=\frac{1}{\mathcal{G}}\prod_{k=1}^{N}\oint\frac{  dz_{k}  z_k^{  -(c+1)  }  }{  2\pi i  }
\sum_{\{ L_{\Ave{ij}=0,1} \}}
\lb \prod_{i=1}^{N}z_{i}^{\sum_{j\neq i} L_{\Ave{ij}}} \rb
\no \\
&&\times \overline{\Tr }
\lsb
 \exp
  \lbb  \beta \sum_{i<j} L_{\Ave{ij}}J_{\Ave{ij}}\sum_{a=1}^{n}x_{i}^{a}x_{j}^{a}-\frac{\beta}{2}u \sum_{i}\sum_{a}(x_{i}^{a})^2   
  \rbb
  \rsb_{J}.
\ee
We hereafter assume the symbols $\overline{\Tr }$ and $\Tr $ denote the integrations over $\{ \V{x}^{a} \}_{a=1}^{n}$ with and without the constraint $\sum_{i}x^{a}_i=N$, respectively. If the argument is specified in a symbol like $\Tr_{y}$, the integration is performed over the variable $y$, not over $\V{x}$. As \Req{graph-trick}, 
\be
&&\sum_{\{ L_{\Ave{ij}}\}=0,1}\prod_{i=1}^{N}z_{i}^{\sum_{j\neq i}L_{\Ave{ij}}}
\lsb
 e^{
  \beta \sum_{i<j} L_{\Ave{ij}}J_{\Ave{ij}}\sum_{a=1}^{n}x_{i}^{a}x_{j}^{a}
 }
  \rsb
\approx 
\prod_{\Ave{ij}}
e^{z_iz_j \lsb e^{\beta J\sum_{a=1}^{n}x_{i}^{a}x_{j}^{a}}   \rsb_{J} }.
\ee
Here we introduce auxiliary variables $\{ y^{a}_1,y^{a}_2 \}_{a=1}^{n}$ and perform the following transformation
\be
\lsb e^{\beta J\sum_{a=1}^{n}x_{i}^{a}x_{j}^{a}}   \rsb_{J} 
=\Tr_{y_1,y_2}
\lsb e^{\beta J\sum_{a=1}^{n}y_{1}^{a}y_{2}^{a}}   \rsb_{J} 
\prod_{a}\delta\lb y^a_1 -x^a_i  \rb 
\delta\lb y^a_2-x^a_j  \rb,
\ee
We introduce an order parameter function $Q\lb y \rb=(1/N)\sum_{i}z_i \prod_{a}\delta\lb y^a-x^a_i  \rb$. The constraint of this relation is expressed by the delta function and the integration over $Q\lb y \rb$. The replica indices of the arguments of these functions are omitted again for simplicity. Employing the Fourier expression of the delta function with auxiliary integrating variables $\Wh{Q}\lb  y \rb$, we get
\be
&& 
\hspace{-10mm}
[Z^n]=\frac{1}{\mathcal{G}^n}\prod_{k=1}^{N}\oint\frac{  dz_{k}  z_k^{  -(c+1)  }  }{  2\pi i  }
\int DQ D\Wh{Q}
 \overline{\Tr }  \exp 
  \Biggl\{
   \Tr_{y_1,y_2} \frac{1}{2}N^2Q(y_1)Q(y_2)\lsb e^{\beta J\sum_{a=1}^{n}y_{1}^{a}y_{2}^{a}}   \rsb_{J} 
 \no \\
 &&
  +\Tr_{y}\Wh{Q}(y)\lb \sum_{i}z_i\prod_{a}\delta(y^a-x^a_i)
  -NQ(y)\rb 
 -\frac{\beta}{2}u\sum_{i}\sum_{a}(x_{i}^{a})^2
 \Biggr\}.
\ee
The symbols $DQ$ and $D\Wh{Q}$ are the integrations over $Q\lb  y \rb$ and $\Wh{Q}\lb  y \rb$ explained above.  The integration over $x$ can now be performed independently over each site 
\be
&&
\overline{\Tr } \prod_{i=1}^{N}e^{\Tr_{y}\Wh{Q}(y)z_i\prod_a\delta(y^a-x^a_i)-\frac{\beta}{2} u \sum_{a} (x_{i}^{a})^2 }
\no \\ &&
=\int \prod_{a=1}^{n}dr_a \prod_{i=1}^{N}
 \lbb 
\lb \prod_{a=1}^{n}  \int_{0}^{\infty} dx_i^{a} \rb
 e^{\sum_{a}r_a(x_{i}^{a}-1)+\Wh{Q}(x_i)z_i-\frac{\beta}{2} u \sum_{a} (x_{i}^{a})^2 }
 \rbb
\no \\
&&
\approx 
\prod_{i=1}^{N}\frac{z_{i}^{c}}{c!}\int \prod_{a=1}^{n}dr_a e^{N\log \Tr_{x} e^{L}  },
\ee
where we put
\be 
 \Tr_{x}e^{L}=\lb \prod_{a=1}^{n}  \int_{0}^{\infty} dx^{a} \rb  e^{\sum_{a}r_a(x^{a}-1)+c \log\Wh{Q}(x)-\frac{\beta}{2} u \sum_{a} (x^{a})^2 }.
\ee
The variables $\{r_a\}_{a=1}^{n}$ are introduced to hold the constraint $\sum_{i}x_i=N$. Summarizing the transformations so far, we get
\be
&&
[Z^n]=\frac{1}{ \mathcal{G}^n } 
\int DQ D\Wh{Q} 
\lb \prod_{a=1}^{n} dr_a\rb
\exp N
 \Biggl\{  
\Tr_{y_1,y_2}\frac{1}{2}NQ(y_1)Q(y_2)
   \lsb e^{  \beta J \sum_{a}y_1^{a}y_2^{a}  } \rsb_{J}
 \no \\ &&
  -\Tr_{y}Q(y)\Wh{Q}(y)
+ \log \Tr_{x} e^{L}-\log c!
  \Biggr\}.
 \Leq{Z^n-pre}
 \ee
Extracting normalization constants from $Q,\Wh{Q}$ as $Q(y)=VP(y),\Wh{Q}(y)=\Wh{V}\Wh{P}(y)$ and taking the saddle-point conditions with respect to $V$ and $\Wh{V}$, we obtain
\be
&&
NV\Tr_{y_1,y_2}P(y_1)P(y_2)
   \lsb e^{  \beta J\sum_{a}y_1^{a}y_2^{a}  } \rsb_{J}=\Wh{V}\Tr_{y}P(y)\Wh{P}(y),
\\
&&c/\Wh{V}=V\Tr_{y}P(y)\Wh{P}(y).
\ee
Inserting these relations into \Req{Z^n-pre}, we see
\be
&&
\phi(n) \equiv \frac{1}{N}\log [Z^n]=
\frac{1}{2}c \log \Tr_{x_1,x_2}P(x_1)P(x_2)
\lsb e^{\beta J \sum_a x_1^{a}x_2^{a}} \rsb_{J}
\no \\ &&
-c\log \Tr_{x}P(x)\Wh{P}(x)+\log \Tr_{x} e^{M}, 
\Leq{phi}
\ee
where
\be 
 \Tr_{x}e^{M}=\lb \prod_{a=1}^{n}  \int_{0}^{\infty} dx ^{a} \rb  e^{\sum_{a}r_a(x^{a}-1)+c_i \log\Wh{P}(x)-\frac{\beta}{2} u \sum_{a} (x^{a})^2 },
\Leq{H_i}
\ee
where we rewrite all the integrating variables as $x$. 
\subsubsection{Replica symmetry}
We assume the replica symmetry (RS):
\be
&&r_a=r,
\\
&&P(x)=\int d\V{\xi}q(\V{\xi})\prod_{a=1}^{n}p(x_a|\V{\xi}),
\\
&&\Wh{P}(x)=\int d\Wh{\V{\xi}}\Wh{q}(\Wh{\V{\xi}})\prod_{a=1}^{n}\Wh{p}(x_a|\Wh{\V{\xi}}).
\Leq{RS}
\ee
Under this assumption, each term of \Req{phi} becomes
\be
&&
\hspace{-0mm}
\Tr_{x_1,x_2}P(x_{1})P(x_{2})\lsb e^{\beta J \sum_{a}x_{1}^{a}x_{2}^{a}}\rsb_{J}
\no \\ &&
\hspace{-0mm}
=
 \int d\V{\xi}_1d\V{\xi}_2 q(\V{\xi}_1)q(\V{\xi}_2) 
  \lsb
   \prod_{a}  \int_{0}^{\infty} dx_1^{a} dx_2^{a}\,
    p(x^{a}_1|\V{\xi}_1)p(x^{a}_2|\V{\xi}_2) 
    e^{\beta J x_{1}^{a}x_{2}^{a}}
  \rsb_{J}
\no \\ &&
\hspace{-0mm}
\equiv
 \int d\V{\xi}_1d\V{\xi}_2 q(\V{\xi}_1)q(\V{\xi}_2) 
  \lsb
  K_{1}^{n}
  \rsb_{J}
\approx 1+n  \int d\V{\xi}_1d\V{\xi}_2 q(\V{\xi}_1)q(\V{\xi}_2) \lsb \log K_{1} \rsb_{J},
\\ &&\hspace{-0mm}
\Tr_{x}P(x)\Wh{P}(x)
 =
  \int d\V{\xi}d\Wh{\V{\xi}} q(\V{\xi})\Wh{q}(\Wh{\V{\xi}}) 
  \prod_{a} \int_{0}^{\infty} dx^a{}
  p(x^{a}|\V{\xi})\Wh{p}(x^{a}_2|\Wh{\V{\xi}})
\no \\ &&\hspace{-0mm}
  \equiv
  \int d\V{\xi}d\Wh{\V{\xi}} q(\V{\xi})\Wh{q}(\Wh{\V{\xi}}) K_2^{n}
 \approx
1+n\int d\V{\xi}d\Wh{\V{\xi}} q(\V{\xi})\Wh{q}(\Wh{\V{\xi}}) \log K_2,
\\
&&
\Tr e^{M} 
=
\int \prod_{l=1}^{c} d\Wh{\V{\xi}}_l q(\Wh{\V{\xi}}_l)
\prod_{a} 
\int_{0}^{\infty}d x^a \, e^{r(x^a-1)-\frac{\beta}{2}u (x^{a})^2} p(x^a|\Wh{\V{\xi}}_1)\cdots p(x^a|\Wh{\V{\xi}}_{c})
\no \\
&&
\equiv
\int \prod_{l=1}^{c} d\Wh{\V{\xi}}_l q(\Wh{\V{\xi}}_l)K_{3}^{n}
\approx 
1+n\int \prod_{l=1}^{c} d\Wh{\V{\xi}}_l q(\Wh{\V{\xi}}_l)\log K_{3},
\ee
Hence,
\be
&&
\hspace{-0mm}
\phi(n)=\frac{1}{2}c\log  \int d\V{\xi}_1d\V{\xi}_2 q(\V{\xi}_1)q(\V{\xi}_2)K_1^n
\no \\ &&
-c\log  \int d\V{\xi}d\Wh{\V{\xi}}q(\V{\xi})\Wh{q}(\Wh{\V{\xi}})K_2^{n} 
+\log \int \prod_{l=1}^{c} d\Wh{\V{\xi}}_l q(\Wh{\V{\xi}}_l)K_{3}^{n}.
\ee
Taking the limit $-\beta f=\lim_{n\to 0}\phi(n)/n$ and extending the integration region with respect to $x$ from $[0;\infty]$ to $[-\infty; \infty]$, we get \Reqss{f-RS}{K_3}. 



\begin{thebibliography}{99}


\bibitem{Motomura:32} Motomura I, {\it On the statistical treatment of communities}, 1932 {\it Zool. Mag., Tokyo} {\bf 44} 379

\bibitem{Corbet:43} Corbet A S, Fisher R A, and Williams C B, {\it The relation between the number of species and the number of individuals in a random sample of an animal population}, 1943  {\it J. Anim. Ecol.} {\bf 12} 42

\bibitem{MacArthur:57} MacArthur R H, {\it On the relative abundance of bird species},  1957 {\it Proc. Natl. Acad. Sci.} {\bf 43} 293

\bibitem{MacArthur:60} MacArthur R H, {\it On the relative abundance of species}, 1960 {\bf Am. Nat.} {\bf 94} 25

\bibitem{Preston:62a} Preston F W, {\it The canonical distribution of commonness and rarity: Part 1}, 1962 {\it Ecology} {\bf 43} 185

\bibitem{Preston:62b} Preston F W, {\it The canonical distribution of commonness and rarity: Part 2}, 1962 {\it Ecology} {\bf 43} 410



\bibitem{Whittaker:70} Whittaker R H, {\it Communities and Ecosystems}, 1970 (Macmillan Publishing Co., New York)

\bibitem{Bazzaz:75} Bazzaz F A, {\it Plant species diversity in oldfield successional ecosystems in southern Illinois}, 1975 {\it Ecology} {\bf 56} 485

\bibitem{May:75} May R M, {\it Patterns of Species Abundance and Diversity}, 1975  (Belknap, Cambridge, pp. 81-120)

\bibitem{Sugihara:80} Sugihara G, {\it Minimal community structure: an explanation of species abundance pattern}, 1980 {\it Am. Nat.} {\bf 116} 770

\bibitem{Nee:91} Nee S, Harvey P H, and May R M, {\it Lifting the veil on abundance patterns}, 1991 {\it Proc. R. Soc. Lond. B} {\bf 243} 161

\bibitem{Tokeshi:99} Tokeshi M, {\it Species Coexistence}, 1999  (Blackwell Science, Oxford)

\bibitem{Hubbell:01} Hubbell S P, {\it The Unified Neutral Theory of Biodiversity and Biogeography}, 2001 (Princeton University Press)

\bibitem{Volkov:03} Volkov I, Banavar J R, Hubbell S P, and Maritan A, {\it Neutral theory and relative species abundance in ecology},  2003 {\it Nature} {\bf 424} 1035

\bibitem{Vallade:03} Vallade M and Houchmandzadeh B, {\it Analytical solution of a neutral model of biodiversity}, 2003 {\it Phys. Rev. E} {\bf 68} 061902

\bibitem{Alonso:04} Alonso D and McKane A J, {\it Sampling Hubbell’s neutral theory of biodiversity}, 2004 {\it Ecology Letters} {\bf 7} 901  

\bibitem{Etienne:05} Etienne R S,  {\it A new sampling theory for neutral biodiversity},  2005 {\it Ecology Letters} {\bf 8} 253

\bibitem{Alonso:06} Alonso D, Etienne R S, and McKane A J, {\it The merits of neutral theory}, 2006 {\it TRENDS in Ecology and Evolution} {\bf 21} 451 

\bibitem{Etienne:07} Etienne R S and Alonso D, {\it Neutral Community Theory: How Stochasticity and Dispersal-Limitation Can Explain Species Coexistence}, 2007 {\it J. Stat. Phys.} {\bf 128} 485

\bibitem{Hofbauer:98} Hofbauer J and Sigmund K, {\it Evolutionary Games and Population Dynamics}, 1998 (Cambridge Univ. Press)

\bibitem{Taylor:78} Taylor P D and Jonker L B, {\it Evolutionary stable strategies and game dynamics}, 1978 {\it Mathematical Bioscience} {\bf 40} 145

\bibitem{Nowak:06} Nowak M A, {\it Evolutionary Dynamics},  2006 (Harvard Univ. Press)

\bibitem{Mougi:12} Mougi A and Kondoh M, {\it Diversity of interaction types and colonial community stability}, 2012 {\it Science} {\bf 337} 349

\bibitem{Eigen:79} Eigen M and Schuster P, {\it The Hypercycle -- A Principle of Natural Self-Organization}, 1979 (Springer)

\bibitem{Ohtsuki:06} Ohtsuki H, Hauert C, Lieberman, E, and Nowak M A, {\it A simple rule for the evolution of cooperation
on graphs and social networks}, 2006 {\it Nature} {\bf 441} 502

\bibitem{Nowak:01} Nowak M A, Komarova N L, and Niyogi P, {\it Evolution of Universal Grammar}, 2001 {\it Science} {\bf 114} 114

\bibitem{Rieger:89} Rieger H, {\it Solvable model of a complex ecosystem with randomly interacting species}, 1989 {\it J. Phys. A: Math. Gen.} {\bf 22} 3447

\bibitem{Diederich:89} Diederich S and Opper M, {\it Replicators with random interactions: A solvable model}, 1989 {\it Phys. Rev. A} {\bf 39} 4333

\bibitem{Oliveira:00} de Oliveira V M and Fontanari J F, {\it Random Replicators with High-Order Interactions}, 2000 {\it Phys. Rev. Lett.} {\bf 85} 4984

\bibitem{Oliveira:01} de Oliveira V M and Fontanari J F, {\it Extinctions in the random replicator model}, 2001 {\it Phys. Rev. E} {\bf 64} 051911

\bibitem{Oliveira:02} de Oliveira V M and Fontanari J F, {\it Complementarity and Diversity in a Soluble Model Ecosystem}, 2002 {\it Phys. Rev. Lett.} {\bf 89} 148101

\bibitem{Oliveira:03} de Oliveira V M, {\it Replicators with Hebb interactions}, 2003 {\it Eur. Phys. J. B} {\bf 31}, 259

\bibitem{Tokita:04} Tokita K, {\it Species Abundance Patterns in Complex Evolutionary Dynamics}, 2004, Phys. Rev. Lett. {\bf 93} 178102

\bibitem{Tokita:06} Tokita K, {\it Statistical mechanics of relative species abundance}, 2006 {\it Ecological Informatics} {\bf 1}, 315

\bibitem{Galla:06} Galla T, {\it Random replicators with asymmetric couplings}, 2006 {\it J. Phys. A: Math. Gen.} {\bf 39} 3853

\bibitem{Yoshino:08} Yoshino Y, Galla T and Tokita K, {\it Rank abundance relations in evolutionary dynamics of random replicators}, 2008 {\it Phys. Rev. E} {\bf 78}, 031924

\bibitem{Galla:12} Galla T, {\it Relative population size, cooperation pressure and strategy correlation in two-population evolutionary dynamics}, 2012 {\it Philosophical Magazine} {\bf 92} 324

\bibitem{May:72} May R M, {\it Will a large complex system be stable?}, 1972 {\it Nature} {\bf 238} 413

\bibitem{Allesina:12} Allesina S and Tang S, {\it Stability criteria for complex ecosystem}, 2012 {\it Nature} {\bf 483} 205

\bibitem{Berlow:99} Berlow E L, {\it Strong effects of weak interactions in ecological communities}, 1999 {\it Nature}, {\bf 398}, 330

\bibitem{Dornelas:08} Dornelas M and Sean R. {\it Connolly Multiple modes in a coral species abundance distribution}, 2008 {\it Ecology Letters} {\bf 11} 1008

\bibitem{Gray:05} Gray J S, Bj\o rges\ae ter  A, and Ugland K I, {\it The impact of rare species on natural assemblages}, 2005 {\it Journal of Animal Ecology}  {\bf 74} 1131

\bibitem{Magurran:03} Magurran  A E and Henderson P A, {\it Explaining the excess of rare species in natural species abundance distributions}, 2003 {\it Nature} 714

\bibitem{Alonso:08} Alonso D, Ostling A, and Etienne R S, {\it The implicit assumption of symmetry and the species abundance distribution}, 2008 {\it Ecology Letters} {\bf 11} 93


\bibitem{STAT} Nishimori H, {\it Statistical Physics of Spin Glasses and Information Processing: An Introduction}, 2001 (Oxford: Oxford University Press)

\bibitem{Kabashima:12} Kabashima Y and Takahashi H, {\it First eigenvalue/eigenvector in sparse random symmetric matrices: influences of degree fluctuation}, 2012 {\it J. Phys. A} {\bf 45} 325001

\bibitem{Obuchi:16}  Obuchi T, Kabashima Y and Tokita K, {\it Multiple peaks of species abundance distributions induced by sparse interactions}, arXiv:1605.09106 







\end{thebibliography}
\end{document}